\documentclass[
aip,
reprint,
supercriptaddress
]{revtex4-2}

\usepackage{siunitx}
\DeclareSIUnit[number-unit-product = \,]{\atpercent}{at.\%} 
\DeclareSIUnit[number-unit-product = \,]{\wtpercent}{wt.\%} 
\DeclareSIUnit[number-unit-product = \,]{\amu}{amu}

\usepackage{graphicx}
\usepackage[outdir=./]{epstopdf}
\usepackage{amsmath, amssymb}
\usepackage{booktabs}
\usepackage{nicefrac}

\usepackage{etoolbox}
\newrobustcmd\B{\DeclareFontSeriesDefault[rm]{bf}{b}\bfseries}

\usepackage{hyperref}

\newcommand{\lwr}[1]{\textsubscript{\protect\raisebox{-1pt}{#1}}}
\newcommand{\upr}[1]{\textsuperscript{\protect\raisebox{1pt}{#1}}}
\newcommand{\mlwr}[1]{_{\mathrm{#1}}}			
\newcommand{\mupr}[1]{^{\mathrm{#1}}}
\newcommand{\uplwr}[2]{\rlap{\upr{#1}}\lwr{#2}}
\newcommand{\muplwr}[2]{\mupr{#1}\mlwr{#2}}

\newcommand{\imag}{\text{i}}

\begin{document}
	
	\title[]{Light absorption and emission by defects in doped nickel oxide}
	
	\author{Robert Karsthof}
	\email{r.m.karsthof@smn.uio.no}
	\affiliation{Centre for Materials Science and Nanotechnology, Universitetet i Oslo, Oslo (Norway)}
	\affiliation{Universit\"{a}t Leipzig, Felix-Bloch-Institut f\"{u}r Festkörperphysik, Leipzig (Germany)}
	
%
	\author{Ymir Kalmann Frodason}
	\affiliation{Department of Physics, Universitetet i Oslo, Oslo (Norway)}

	\author{Augustinas Galeckas}
	\affiliation{Department of Physics, Universitetet i Oslo, Oslo (Norway)}
	
	\author{Philip Michael Weiser}
	\affiliation{Centre for Materials Science and Nanotechnology, Universitetet i Oslo, Oslo (Norway)}
	
%
	
	\author{Vitaly Zviagin}
	\affiliation{Universit\"{a}t Leipzig, Felix-Bloch-Institut f\"{u}r Festkörperphysik, Leipzig (Germany)}

	\author{Marius Grundmann}
	\affiliation{Universit\"{a}t Leipzig, Felix-Bloch-Institut f\"{u}r Festkörperphysik, Leipzig (Germany)}
	

	\date{\today}
	
\begin{abstract}
	Nickel oxide is a versatile $p$-type semiconducting oxide with many applications in opto-electronic devices, but high doping concentrations are often required to achieve necessary electrical conductivity. In contrast to many other transparent oxide semiconductors, even moderate levels of doping of NiO can lead to significant optical absorption in the visible spectral range, limiting the application range of the material. This correlation has been reported extensively in literature, but its origin has been unknown until now. This work combines experimental data on optical properties from a variety of NiO samples with results from hybrid density functional theory calculations. It shows that strong electron-phonon interaction leads to a significant blue shift (\SIrange[range-phrase=-,range-units=single]{0.6}{1}{\electronvolt}) of electronic transitions from the valence band maximum to defect states by light absorption with respect to the thermodynamic charge transition levels. This essentially renders NiO a narrow-gap semiconductor by defect band formation already at moderate doping levels, with strong light absorption for photon energies of approximately \SI{1}{\electronvolt}. The calculations are also shown to be fully consistent with experimental data on defect-related light emission in NiO.
\end{abstract}

\maketitle

\section{Introduction}
	
Nickel oxide is one of the few examples of wide-gap semiconducting metal oxides with holes as majority carriers. Combined with its high optical transparency in its pure, defect-free form, and its comparably high-lying valence and conduction bands, it has become widely used in various opto-electronic devices as electron-blocking, hole-conducting top electrode where transmitting visible light is critical for device functionality. The most prominent example is perovskite solar cells where NiO has been widely explored as a hole transport layer \cite{Xu2019,Li2019,Feng2020}.
To achieve at least moderately conductive NiO films, doping by monovalent ions such as Li, Na, K, Cu or Ag can be employed \cite{Bielanski1962,Lany2007,Yang2012,Zhang2018,Egbo2020}. Alternatively, inducing an excess of oxygen/ deficiency of nickel leads to the formation of nickel vacancies V\lwr{Ni} that act as acceptors, albeit not producing the same level of conductivity as the first method. In contrast to ``conventional'' wide-gap semiconductors, such as ZnO or In\lwr{2}O\lwr{3}, where high optical transmittance in the visible spectral range is retained also for carrier concentrations in the degenerate regime, doping-induced optical absorbance in NiO is pronounced, leading to a reduction of average visible transmittance $T$ of the order of \SI{50}{\percent} already at hole concentrations around \SI{e18}{\per\cubic\centi\meter} \cite{Egbo2020,Egbo2020a}. From a device fabrication perspective, this makes it necessary to accept a trade-off between the optical and electrical performance of NiO electrodes, as both high $T$ and $\sigma$ cannot be achieved at the same time. 
In a recent publication \cite{Karsthof2020}, we have determined the $(0/-)$ charge transition levels (CTLs) of the Li\lwr{Ni} and V\lwr{Ni} acceptors as \SI{0.19}{\electronvolt} and \SI{0.41}{\electronvolt} above the valence band maximum (VBM), respectively, i.e. shallow enough to be at least partially ionizable at room temperature.  
In this publication, we demonstrate that doping NiO with Li or with Ni vacancies leads to the formation of detectable defect bands in the band gap that enable optical transitions in the near infrared spectral range (\SIrange{1}{2}{\electronvolt}). 
Based on hybrid functional calculations, we construct configuration coordinate diagrams describing optical charge-state transitions involving the Li\lwr{Ni} and V\lwr{Ni} acceptors. Despite their relatively small ionization energies, these acceptors are predicted to exhibit polaronic localized hole states, resulting in strong electron-phonon coupling and optical transitions with significant Franck-Condon shifts. This leads to a large blue shift of the electronic transitions induced by photon absorption when compared to the thermodynamic charge transition levels, as they have been observed in Ref.~\cite{Karsthof2020}. Additionally, the hybrid-functional calculations suggest photon emission by charge carrier-defect recombination with emission spectra that agree well with experimental photoluminescence data.

\section{Methods}

\subsection{Materials}

Various NiO specimen were studied. A $\SI{5}{\milli\meter}\times\SI{5}{\milli\meter}\times\SI{1}{\milli\meter}$ (100)-oriented as-cut (unpolished) NiO single crystal was purchased from MaTecK GmbH (Germany). This mirror furnace-grown crystal was black in color, fully opaque, and electrically insulating. NiO thin films were deposited by means of pulsed laser deposition (PLD) and reactive DC magnetron sputtering. For the PLD-grown films, (100) oriented MgO substrates were used. For PLD growth, pellets of either pure NiO (\SI{99.998}{\percent} purity) or Li-doped NiO (admixture of \SI{10}{\wtpercent} Li\lwr{2}CO\lwr{3}) were fabricated by mixing and ball milling of the powders, pressing at \SI{7}{\bar} into \SI{26}{\milli\meter}-diameter disks in a hydraulic press, and sintering the disks for \SI{24}{\hour} at \SI{1350}{\degreeCelsius}. PLD was carried out in a home-built vacuum chamber with a KrF excimer laser ($\lambda = \SI{248}{\nano\meter}$, pulse energy \SI{650}{\milli\joule}, repetition rate \SI{10}{\hertz}). The substrate temperature during growth was adjusted to \SI{300}{\degreeCelsius}. Oxygen was used as background gas, and the O\lwr{2} partial pressure was \SI{0.01}{\milli\bar}. Films resulting from PLD growth were either insulating when grown from the pure NiO pellet, or conductive and visible light-absorbing/ black when Li doping was employed, and will be referred to as \textit{pure NiO} and \textit{Li-doped NiO}, respectively, in the further course of this paper.

DC magnetron sputtering was performed in a reactive Ar/O\lwr{2} atmosphere from a metallic Ni target on fused silica substrates (target-substrate distance \SI{40}{\milli\meter}), with $p\mlwr{O2}$ varying between \SI{1.1e-3}{\milli\bar} and \SI{2.8e-2}{\milli\bar} (while the Ar partial pressure was kept constant at \SI{1.8e-2}{\milli\bar} to achieve variable Ni:O ratios. DC sputtering power was set to \SI{30}{\watt}. No substrate heating was employed. Before film deposition, the target was pre-sputtered \textit{in situ} for \SI{10}{\minute} in pure Ar ($p\mlwr{Ar} = \SI{3.2e-2}{\milli\bar}$). Because of the resulting intrinsic doping with the V\lwr{Ni} acceptor, these films will be referred to as \textit{V\lwr{Ni}-doped NiO}. They are slightly absorbing/ semi-transparent depending on $p\mlwr{O2}$ used, and of gray color.

\subsection{Thin film characterization}

X-ray diffraction was recorded with a Philips X'pert diffractometer using Cu K$\alpha$ radiation ($\lambda = \SI{1.5406}{\angstrom}$). Optical transmittance and (direct) reflectance measurements were performed with Perkin Lambda 19 UV/VIS/NIR and Shimadzu SolidSpec-3700 DUV spectrophotometers, using tungsten, halogen, and deuterium light sources. The thin film complex UV dielectric function was determined using a variable-angle spectroscopic ellipsometry (J.A. Woollam Co.) in a polarizer-compensator-sample-analyzer configuration. Measurements were carried out at ambient conditions in the spectral range from \SI{0.5}{\electronvolt} to \SI{8.5}{\electronvolt}8.5 eV. 
For the visible spectral range, spectroscopic ellipsometry was carried out using a M-2000 ellipsometer (J.A. Woolam Co.). Evaluation and modeling of the spectra was done with the Woollam CompleteEASE software. \\

The photoluminescence (PL) measurements were carried out at \SI{10}{\kelvin} using a \SI{325}{\nano\meter} wavelength of cw-HeCd laser (photon energy \SI{3.81}{\electronvolt}) and a \SI{246}{\nano\meter} (\SI{5.04}{\electronvolt}) line of a frequency-tripled Ti:sapphire laser for excitation with a power density of \SI{20}{\watt\per\square\centi\meter} in both cases. The PL emission was analyzed by a ﬁber-optic (Ocean Optics, usb4000) and imaging spectrometer systems (Horiba iHR320 coupled to Andor iXon888 EMCCD) with spectral resolution below \SI{2}{\nano\meter}.

\subsection{Computational methods}

\begin{figure}
	\centering
	\includegraphics[width=\columnwidth]{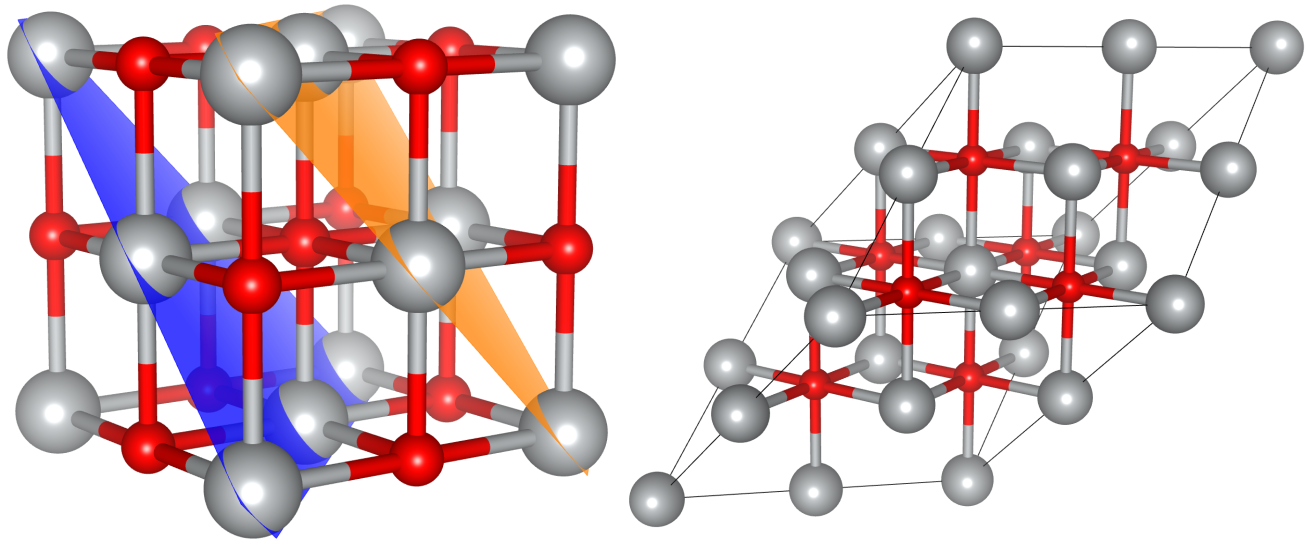}
	\caption{Left: conventional unit cell of NiO, showing the AFII ferromagnetic ordering, as indicated by the blue and orange (111) lattice planes. Right: $2\times2\times2$ repetition of the rhombohedral primitive NiO unit cell, used for bulk calculations.}
	\label{fig:unit_cell}
\end{figure}

All first-principles calculations were based on the generalized Kohn-Sham theory with the projector-augmented wave method \cite{Bloechl1994,Kresse1999}, as implemented in VASP \cite{Kresse1996}. The Ni $3d$ electrons were treated explicitly as valence electrons. In order to obtain the energetically preferred antiferromagnetic spin ordering along the [111] direction in NiO (AFII phase), shown in Fig.~\ref{fig:unit_cell}, bulk calculations were performed using a rhombohedral unit cell with eight Ni and eight O atoms ($2\times2\times2$ repetition of the primitive unit cell). For these calculations, the energy cutoff was set to \SI{520}{\electronvolt}, and the Brillouin Zone was sampled by a $\Gamma$-centered $3\times3\times3$ grid of $k$-points.

Liu et al. \cite{Liu2019} recently performed an assessment of several different hybrid functionals in describing the challenging antiferromagnetic transition-metal monoxides MnO, FeO, CoO, and NiO, which are considered to be prototypical strongly correlated electron systems. Based on the agreement with available experimental data (fundamental band gap, magnetic moments, valence band density of states, and ion-clamped macroscopic dielectric constant) in that study, the HSE03 functional is employed here \cite{Krukau2006}, meaning that the fraction of screened Fock exchange $\alpha$ is set to \num{0.25}, and the range-separation parameter $\mu$ is set to \num{0.30}. The resulting lattice parameter and indirect band gap value are listed in Table~\ref{tab:NiO_DFT_calc_param}, along with experimental data. All defect calculations were done using rhombohedral 128-atom supercells ($4\times4\times4$ repetition of the primitive unit cell), a plane-wave energy cutoff of \SI{520}{\electronvolt}, and a single special $k$-point at (0.25, 0.25, 0.25). Defect formation energies and thermodynamic charge-state transition levels were calculated by following the well-established formalism described in Ref.~\cite{Freysoldt2014}. For example, the formation energy of Li\uplwr{\textit{q}}{Ni} in charge state $q$ is given by

\begin{equation}
	E\mupr{\text{f}}[\text{Li}\muplwr{q}{\text{Ni}}] = E\mlwr{tot}[\text{Li}\muplwr{q}{\text{Ni}}] - E\mlwr{tot}[\text{bulk}] + \mu\mlwr{\text{Ni}} - \mu\mlwr{\text{Li}} + qE\mlwr{F},
	\label{eq:formation_energy}
\end{equation}

where $E\mlwr{tot}[\text{Li}\muplwr{q}{\text{Ni}}]$ and $E\mlwr{tot}[\text{bulk}]$ are the total energies of the defect-containing and pristine supercells, $\mu\mlwr{\text{Ni}}$ and $\mu\mlwr{\text{Li}}$ are the chemical potentials of Ni and Li, and $E\mlwr{F}$ is the Fermi-level position relative to the bulk VBM. Upper bounds on the chemical potential of Ni and O are given by the energy of fcc metal Ni and the O$_2$ molecule, corresponding to Ni-rich and O-rich limits, respectively. These upper limits impose lower bounds on the corresponding other species, given by the thermodynamic stability condition $\Delta\mu\mlwr{\text{Ni}} + \Delta\mu\mlwr{\text{Ni}} = \Delta H\mlwr{f}(\text{NiO})$, where $\Delta H\mlwr{f}(\text{NiO})$ is the formation enthalpy per formula unit of NiO. $\mu\mlwr{\text{Li}}$ was referenced to the solubility-limiting phase Li$_2$O in both the Ni-rich and O-rich limits. For charged defects, the total energies were corrected using the schemes outlined in Refs.~\cite{Freysoldt2009,Kumagai2014,Gake2020}, with the static (ion-clamped) dielectric constant of \num{11.6} (\num{5.13}) \cite{Gielisse1965}. Optical absorption and emission energies were estimated using the one-dimensional configuration coordinate (CC) model, using model parameters obtained from the hybrid functional calculations, as described in Refs.~\cite{Alkauskas2012,Alkauskas2016}. For the small hole polaron, the self-trapping energy and optical emission energy was calculated as explained in Ref.~\cite{Varley2012}.


\begin{table}
	\caption{Lattice parameters and indirect band gap $E\muplwr{i}{g}$. Experimental values are listed for comparison. See Ref.~\cite{Liu2019} for comparison of other bulk parameters of NiO.}
	\label{tab:NiO_DFT_calc_param}
	\begin{tabular}{@{}lS[table-format=1.2]cS[table-format=1.2]c@{}}
	\toprule
		&	{$a$ (\si{\angstrom})}	& ref. &	{$E\muplwr{i}{g}$ (\si{\electronvolt})} & ref.	\\
	\midrule
	HSE03	& 4.168 &	& 4.28 & \\
	experiment	& 4.171 &\cite{Carey1991} & 4.0 & \cite{Kurmaev2008} \\
			& 	&		&	4.3 &  \cite{Sawatzky1984} \\
	\bottomrule
	\end{tabular}
\end{table}


\section{Results and Discussion}

\subsection{Structural properties}

\begin{figure}
	\centering
	\includegraphics[width=\columnwidth]{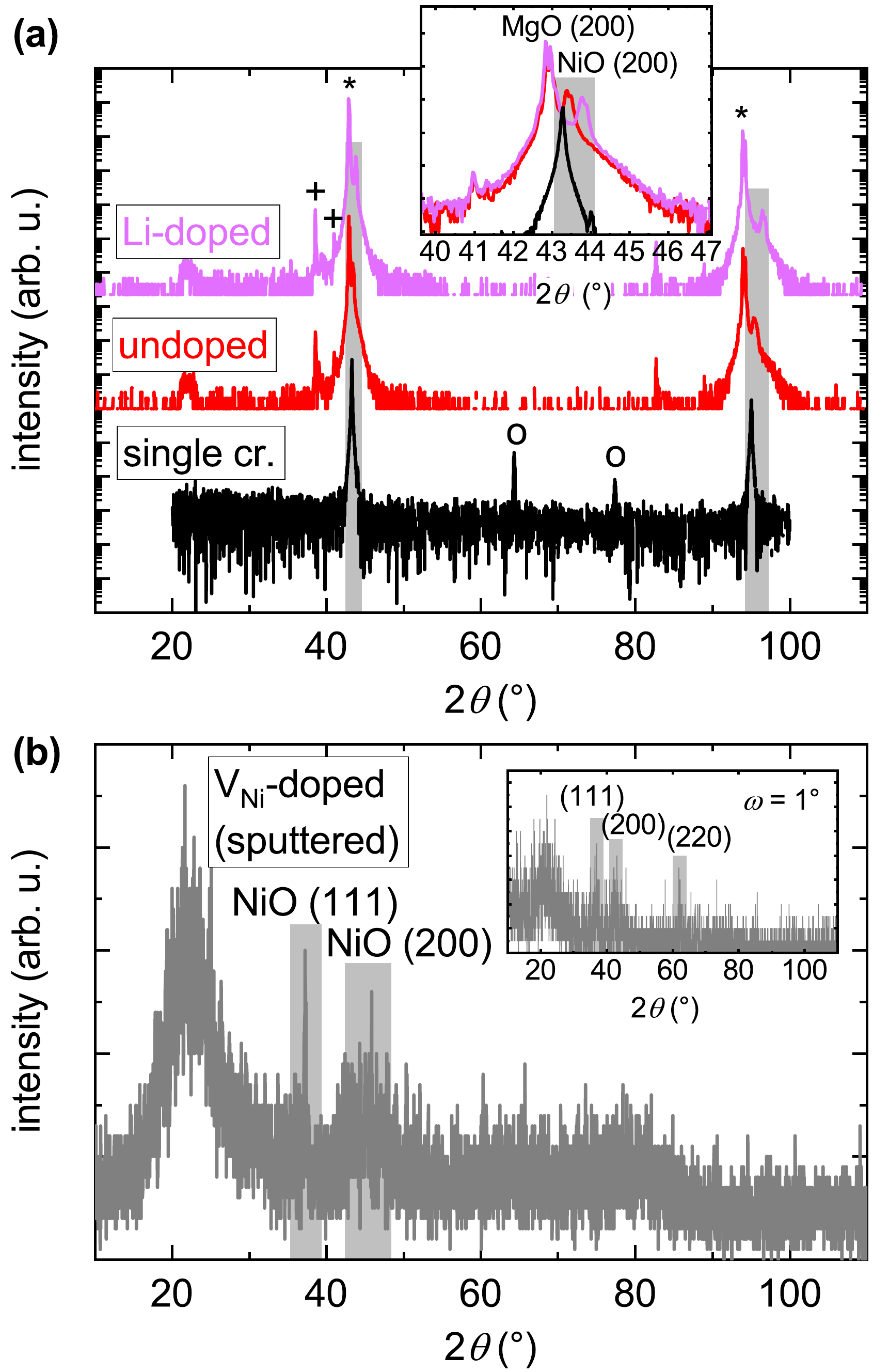}
	\caption{X-ray diffraction patterns of different NiO samples. \textbf{(a)} Single crystal, as well as undoped and Li-doped doped NiO films grown by PLD on MgO substrates. (*) MgO-related reflexes, (+) secondary reflexes due to Cu K-$\beta$ and W K-$\alpha$ spectral lines, (o) reflexes from the sample holder. Inset shows enlarged version of the angular region around the (200) reflexes. \textbf{(b)} DC magnetron-sputtered films on fused silica. Inset: diffractograph obtained under grazing incidence ($\omega = \SI{1}{\degree}$).}
	\label{fig:XRD_all_samples}
\end{figure}

Fig.~\ref{fig:XRD_all_samples} shows the results of the structural investigation by X-ray diffraction. The NiO single crystal exhibits narrow reflexes related to the (100) lattice planes. The (200) lattice plane spacing amounts to $d\mlwr{200} = \SI{2.09}{\angstrom}$, corresponding to an out-of-plane lattice constant of \SI{4.18}{\angstrom}, close to the bulk value ($a\muplwr{NiO}{0,ref} = \SI{4.17}{\angstrom}$). 
Also the epitaxial films (pure and Li-doped) grown by PLD (Fig.~\ref{fig:XRD_all_samples}(a)) are (100) oriented, without secondary film reflexes. The out-of-plane lattice constant of the undoped sample amounts to $a\muplwr{NiO}{0} = \SI{4.168}{\angstrom}$ which is only marginally lower than the reported bulk value, which could be because of slight lattice compression correlated with the in-plane tensile strain induced by the MgO substrate with sightly larger ($a\uplwr{MgO}{0,ref} = \SI{4.21}{\angstrom}$). In the Li-doped sample, the out-of-plane lattice constant is $a\muplwr{NiO}{0} = \SI{4.132}{\angstrom}$ which, based on the reference value for Ni\lwr{0.7}Li\lwr{0.3}O, and using Vegard's law, gives a Li content $x = \num{0.21}$. 

In Fig.~\ref{fig:XRD_all_samples}(b), diffractograms of a reactively sputtered NiO film are shown, exhibiting (besides the broad signal at \SI{22}{\degree} originating from the amorphous glass substrate) very weak NiO (111) and (200) reflexes. When a grazing incidence geometry is used (inset of Fig.~\ref{fig:XRD_all_samples}(b)), these become slightly better visible, and also the (220) reflex can be discerned. The fact that the signal is barely above the noise level can be used to infer that the grains in these polycrystalline films have dimensions of only a few \si{\nano\meter}.

\subsection{Electrical properties}

The electrical conductivity of the studied samples was determined by four-point probe measurements. The values are listed in Table~\ref{tab:electrical_prop}. Even though the single crystal is fully opaque, it has the second-lowest conductivity of all the samples. For the sputtered films, it can be noted that the conductivity decreases with increasing oxygen partial pressure during deposition, in contrast to the typically observed behavior (higher $p\mlwr{O2}$ causes lower formation energy for V\lwr{Ni} acceptor defects). This is likely caused by target poisoning, i.e. a high oxygen supply has lead to an oxidation of the metallic Ni target surface before target ablation. The composition of the resulting film in this case is closer to stoichiometry for higher $p\mlwr{O2}$ than for a lower one.

To determine the carrier type, the hot-probe technique was used. $p$-type conductivity was found on all samples except the undoped thin film on MgO, where no thermovoltage could be detected.

\begin{table*}
	\caption{Electrical conductivity $\sigma$ of the samples under study.}
	\label{tab:electrical_prop}
	\begin{tabular}{cccccccc}
	\toprule
	 & \multicolumn{2}{c}{epitaxial} & & \multicolumn{4}{c}{sputtered ($p\mlwr{O2}$/ \si{\milli\bar})} \\
	 \cmidrule{2-3} \cmidrule{5-8}
	 sample & undoped & Li-doped & single crystal & \num{1.1e-3} & \num{3.5e-3} & \num{1.75e-2} & \num{2.8e-2} \\
	 \midrule	
	$\sigma$ (\si{\siemens\per\meter}) &  \num{2.5e-5} & \num{500} & \num{2.4e-6} & \num{380} & \num{120} & \num{20} & \num{12} \\
	\bottomrule
	\end{tabular}
\end{table*}

\subsection{Optical absorption}

\begin{figure}
	\centering
	\includegraphics[width=\columnwidth]{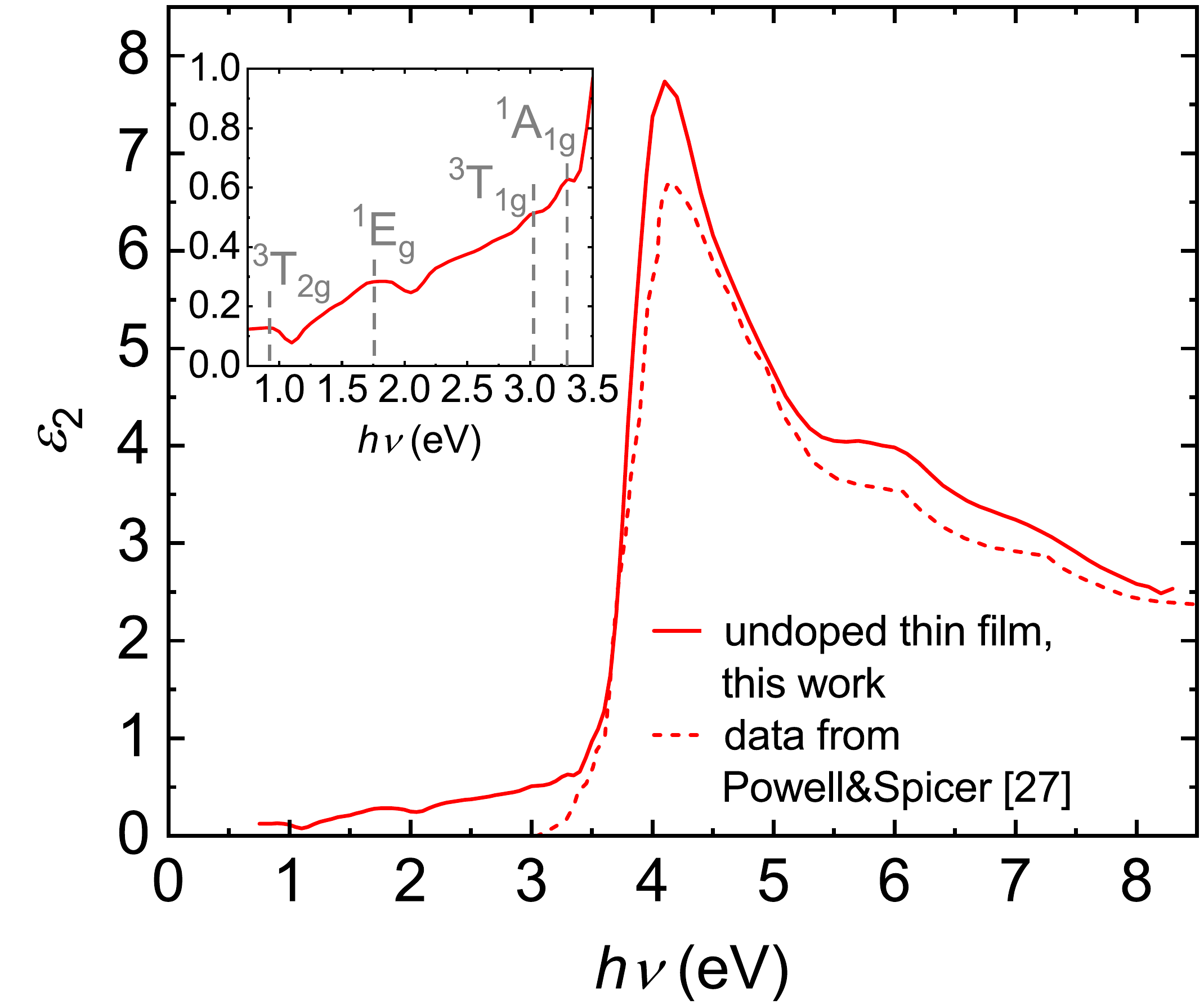}
	\caption{Imaginary part $\varepsilon\mlwr{2}$ of the dielectric function $\varepsilon\mupr{*}$ for an undoped NiO layer grown on MgO substrate, obtained from spectroscopic ellipsometry. Red dashed line shows data from Powell and Spicer \cite{Powell1970}, extracted from reflecometry measurements. Inset: enlarged Version of near-IR and visible spectral ranges, with the main absorption lines from Ni\upr{2+} ions highlighted.}
	\label{fig:pure_NiO_eps2}
\end{figure}

In Fig.~\ref{fig:pure_NiO_eps2}, the imaginary component $\varepsilon_2$ of the complex dielectric function $\varepsilon\mupr{*} = \varepsilon_1 + \imag \varepsilon_2$ is shown for an undoped, epitaxial NiO thin film grown on MgO substrate. The absorption edge is well visible around \SI{4}{\electronvolt} where $\varepsilon_2$ reaches a maximum. Reference data from Ref.~\cite{Powell1970} is also shown, exhibiting the same features. The origin of the transitions observed in the range 4-\SI{7}{\electronvolt} has been ascribed to a series of $p$-$d$ charge transfer excitations \cite{Sokolov2012}. The inset of Fig.~\ref{fig:pure_NiO_eps2} shows the near-infrared and visible spectral range where $\varepsilon_2$ is slowly increasing, with some weak but visible features superimposed that have first been described by Reinen \cite{Reinen1965} and assigned to internal optical transitions on Ni\upr{2+} ions -- $d$-$d$ transitions that are forbidden by selection rules, but obtain non-zero transition matrix elements from disorder, mostly by defects and phonons. These lines are centered mainly in the red and blue ranges, and therefore can give otherwise optically clean samples a light-green hue. The reference data from Ref.~\cite{Powell1970} does not contain these lines, likely because a total reflectance measurement was used which may not be sensitive enough to detect the weak electronic transitions on the Ni\upr{2+} ions.
%

\begin{figure}
	\centering
	\includegraphics[width=\columnwidth]{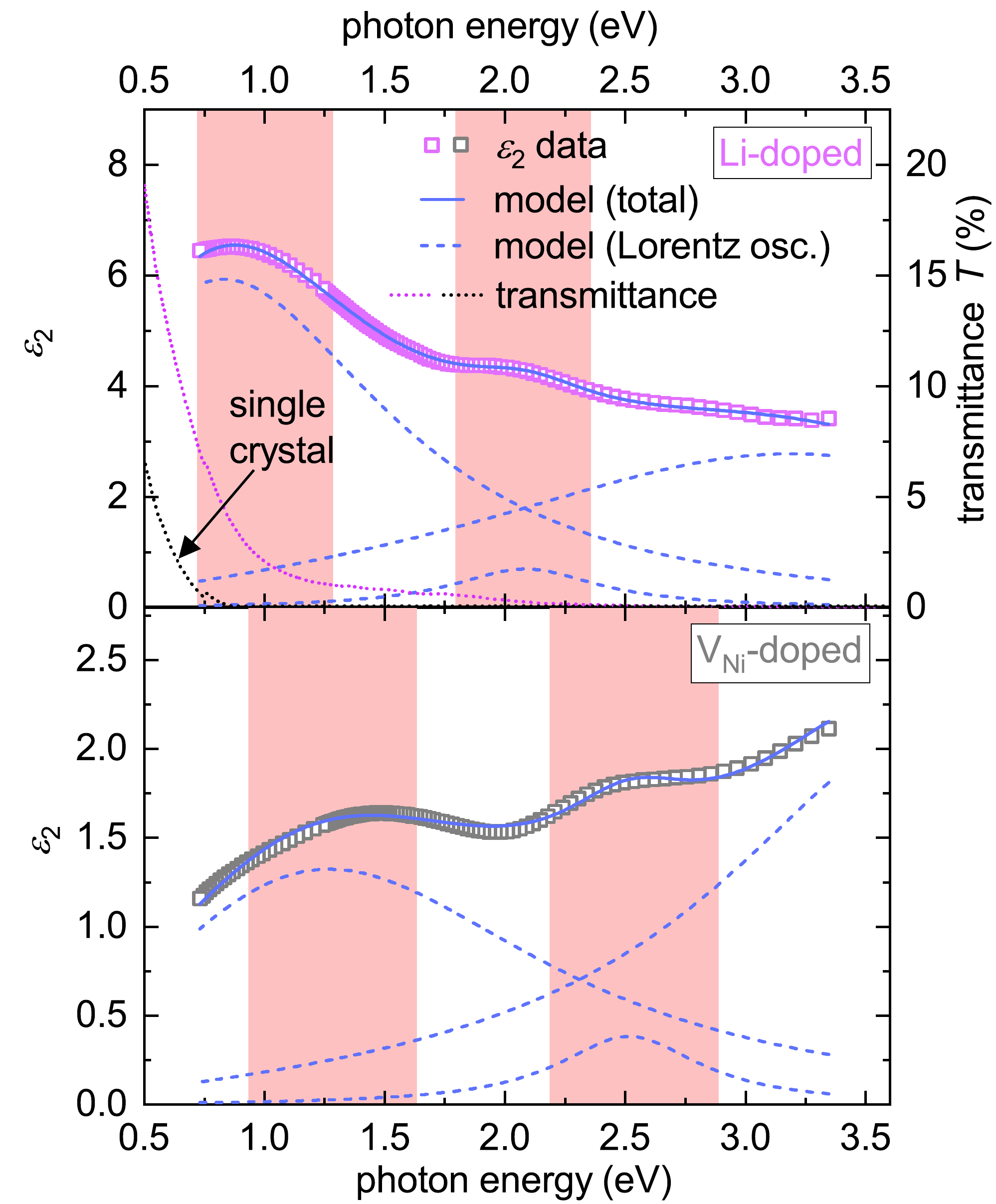}
	\caption{$\varepsilon_2$ data from Li-doped (top) and V\lwr{Ni}-doped (bottom) NiO films obtained by numerical modeling of ellipsometry data, as well as from fitting by a parametrized model containing three Lorentz oscillators. The two absorption bands are highlighted in red. Dotted lines in (a) show increased optical transmittance for energies below approximately \SI{1}{\electronvolt} for the Li-doped thin film and the single crystal. The V\lwr{Ni}-doped sample was grown at a O\lwr{2} partial pressure of \SI{1.75e-2}{\milli\bar}.}
	\label{fig:Li+VNi-doped_eps2_modelling}
\end{figure}
	
We study next the induced changes of the optical absorption properties upon Li and V\lwr{Ni}. 
The imaginary part of the dielectric function $\varepsilon_2$ of a Li-doped and a V\lwr{Ni}-doped (sputtered) NiO thin film is determined from numerical modeling of ellipsometry data, including a fit of this numerical dataset using a parameterized model (see Fig.~\ref{fig:Li+VNi-doped_eps2_modelling}). The formation of two absorption bands is well documented for both doping methods, centered at approximately \SI{1}{\electronvolt} and \SI{2}{\electronvolt} for the Li-doped and \SI{1.25}{\electronvolt} and \SI{2.5}{\electronvolt} for the V\lwr{Ni}-doped (here: $p\mlwr{O2} = \SI{1.75e-2}{\milli\bar}$) case, respectively. The parameterized model contains three Lorentz oscillators (two for the defect bands, one for the band-to-band transition at \SI{4}{\electronvolt}) to reproduce the optical transitions in the near-IR and visible spectral ranges. 
Table~\ref{tab:param_eps2_modeling} summarizes the parameter values for all three oscillators and the two dopant types. 

For the Li-doped sample, which is visibly opaque, Fig.~\ref{fig:Li+VNi-doped_eps2_modelling} shows that the optical transmittance $T$ below \SI{1}{\electronvolt} increases again, indicating that this sample resembles a narrow-band gap material. This interpretation is also supported by the maximum value of $\varepsilon_2 \approx \num{6.5}$ at \SI{1}{\electronvolt}, which is almost as large as that at the original absorption edge in the undoped sample in Fig.~\ref{fig:pure_NiO_eps2} ($\varepsilon_2 = \num{7.5}$). Such high oscillator strengths of electronic transitions are usually found for band-to-band transitions. A similar transmission window for low photon energies is also seen in the single crystal (which was manually polished by SiC and diamond paper to increase the signal quality for the optical transmission measurement), also displayed in the top panel of Fig.~\ref{fig:Li+VNi-doped_eps2_modelling}. This sample is shown to become transparent below photon energies of around \SI{0.8}{\electronvolt}, indicating the existence of one or more absorption bands similar to the one observed for the doped thin films. The fact that the single crystal shows insulating behavior, in contrast to the Li-doped film, can be interpreted as evidence for hole compensation by donor states. These may originate from unknown impurities contained in the sample. Due to the roughness of the unpolished single crystal, ellipsometric measurements were not possible on this specimen.
For the V\lwr{Ni}-doped case, the absorption bands are shifted to higher photon energies, and are slightly broader than in the Li-doped sample. The doping-induced oscillator strengths are lower in this sample than in the Li-doped one, which corresponds to the difference in their electrical conductivity ($\sigma\mlwr{Li} = \SI{500}{\siemens\per\meter}$, $\sigma\mlwr{VNi} = \SI{20}{\siemens\per\meter}$).

\begin{table*}
	\caption{Parameters of the Lorentz oscillator model used for fitting the $\varepsilon_2$ spectra from Fig.~\ref{fig:Li+VNi-doped_eps2_modelling}.}
	\label{tab:param_eps2_modeling}
	\sisetup{detect-weight,
	         mode=text, 
	         table-format=2.0
	         }
	\begin{tabular}{cS[table-format=1.2]S[table-format=1.2]S[table-format=1.2]S[table-format=1.2]S[table-format=1.2]S[table-format=1.2]S[table-format=1.2]S[table-format=1.2]S[table-format=1.2]}
	\toprule
	sample & \multicolumn{3}{c}{{Lorentz 1}} & \multicolumn{3}{c}{{Lorentz 2}} & \multicolumn{3}{c}{{Lorentz 3}} \\
			& {$\Delta \varepsilon\mupr{1}$} & {$\beta\mupr{1}$ (\si{\electronvolt})} & {$E\muplwr{1}{t}$ (\si{\electronvolt})} & {$\Delta \varepsilon\mupr{2}$} & {$\beta\mupr{2}$ (\si{\electronvolt})} & {$E\muplwr{2}{t}$ (\si{\electronvolt})} &{$\Delta \varepsilon\mupr{3}$} & {$\beta\mupr{3}$ (\si{\electronvolt})} & {$E\muplwr{3}{t}$ (\si{\electronvolt})} \\
	\cmidrule(l{0.25em}r{0.25em}){2-4} \cmidrule(l{0.25em}r{0.25em}){5-7} \cmidrule(l{0.25em}r{0.25em}){8-10}
	Li-doped & \num{5.04} & \num{1.86} & \B \num{1.18} & \num{0.69} & \num{0.75} & \num{2.12} & \num{2.63} & \num{3.26} & \num{3.56} \\
	V\lwr{Ni}-doped & \num{1.16} & \num{2.42} & \B \num{1.68} & \num{0.38} & \num{0.72} & \num{2.54} & \num{2.09} & \num{2.21} & \num{3.96} \\
	\bottomrule
	\end{tabular}
\end{table*}

\begin{figure}
	\centering
	\includegraphics[width=\columnwidth]{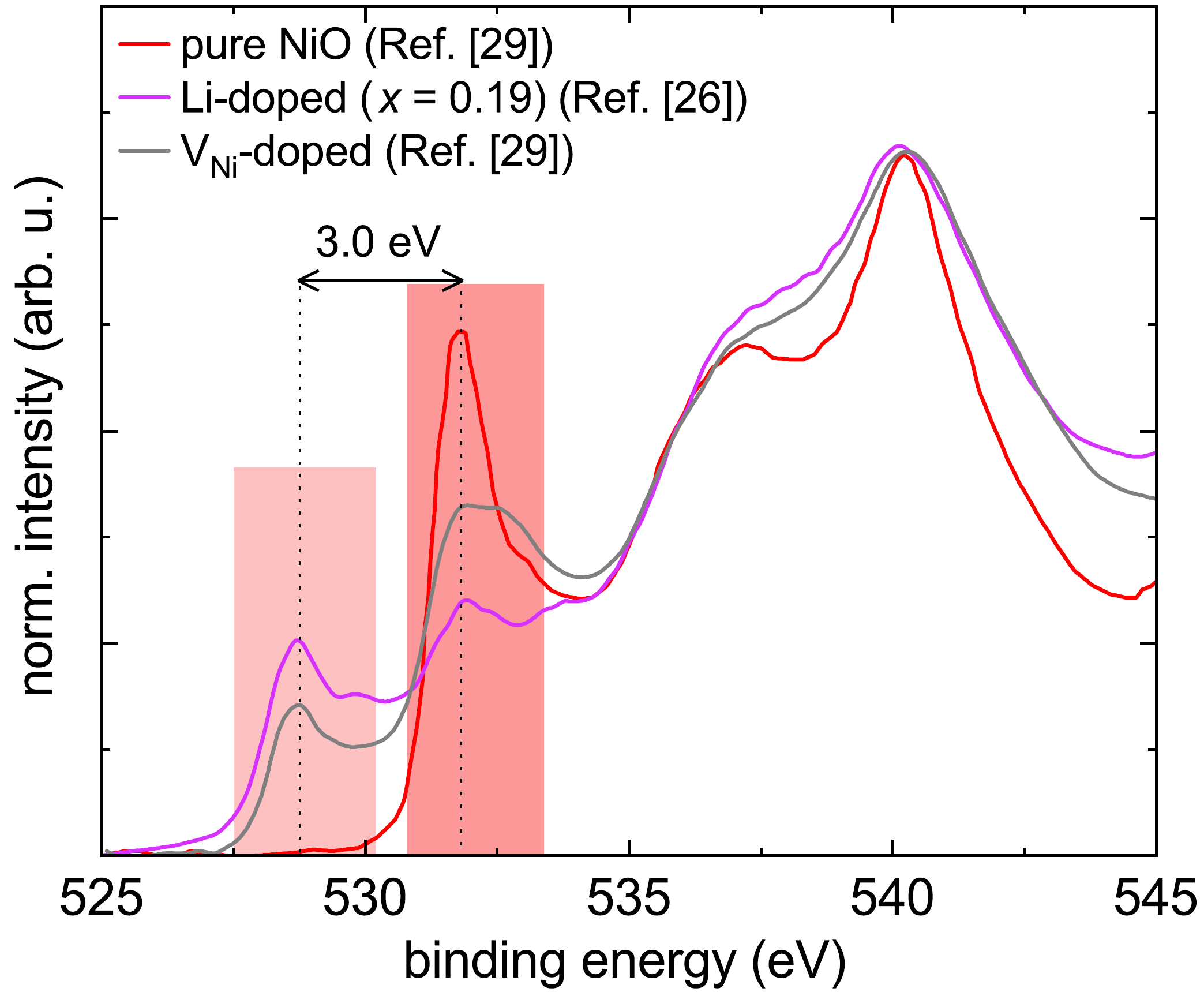}
	\caption{X-ray absorption spectra at the O K-edge of NiO for pure, Li-doped and V\lwr{Ni}-doped NiO. Data extracted from Refs.~\cite{Kuiper1989,Mossanek2013}. Colored areas are spectral ranges exhibiting excitation from O $1s$ into defect-related (lighter red) and conduction band (darker red) states. The intensities have been normalized such that the curves coincide at their maxima at \SI{540}{\electronvolt}.}
	\label{fig:XAS_O_K-edge}
\end{figure}

The two Li doping-induced absorption bands observed in this work have been described in literature before \cite{Kuiper1989,Reinert1995,Zhang2018}. The authors of Ref.~\cite{Zhang2018} suggest an energy scheme of the electronic states where Li doping leads to the formation of a set of broadened in-gap states approximately \SI{1.1}{\electronvolt} above the valence band maximum (VBM). 
The excitation from the VBM and a deeper-lying valence band likely related to $O 2p$ states into the Li acceptor band is assigned as the origin for the two absorption bands. This picture has been supported by experiments on Li-doped NiO probing the electronic structure of the material, such as x-ray absorption spectroscopy (XAS) \cite{Kuiper1989}, and electron energy loss spectroscopy (EELS) \cite{Reinert1995}. 
Both methods show that upon Li incorporation, a new set of electronic states emerges approximately 
\SI{1.2}{\electronvolt} above the valence band edge. A characteristic feature of the emergence of this band is \textit{spectral weight redistribution}: the replacement of a magnetic Ni ion (single spin-degenerate state) by a non-magnetic Li ion (double-degenerate state) creates the in-gap states with twice the rate by which conduction band states are being removed. This leads to a fast decrease of CB-related features. In Fig.~\ref{fig:XAS_O_K-edge}, O $K$edge XAS spectra extracted from Refs.~\cite{Kuiper1989,Mossanek2013} are shown, obtained on undoped, Li-doped, and V\lwr{Ni}-doped material. The formation of the doping-induced band approximately \SI{3}{\electronvolt} below the CBM-related feature at \SI{532}{\electronvolt} is clearly visible in both cases. Simultaneously, the shape and intensity of the \SI{532}{\electronvolt} peak changes drastically because of spectral weight redistribution.  
It is interesting to note that the peak at \SI{528}{\electronvolt} seen in doped samples is often claimed to be due to Ni\upr{3+} states, such as in Ref.~\cite{Mossanek2013} which refers to Kuiper \textit{et al.} \cite{Kuiper1989}. Therein, however, it is stated that the large exchange interaction of the extra hole introduced by doping with the surrounding Ni\upr{2+} spins only makes it appear as if there was Ni\upr{3+} present. The authors conjecture that the \SI{528}{\electronvolt} peak, in fact, originates from ``impurity states in the gap, most likely bound to the effectively negative charged lithium sites'', and therefore support the defect interpretation. 
The Ni\upr{3+} picture can be thought to describe the situation from the perspective of the intact Ni sublattice, whereas the defect picture describes the deviation from the perfect lattice periodicity. However, according to Ref.~\cite{Kuiper1989}, the idea of actual Ni\upr{3+} ions lacks experimental support.

\begin{figure}
	\centering
	\includegraphics[width=\columnwidth]{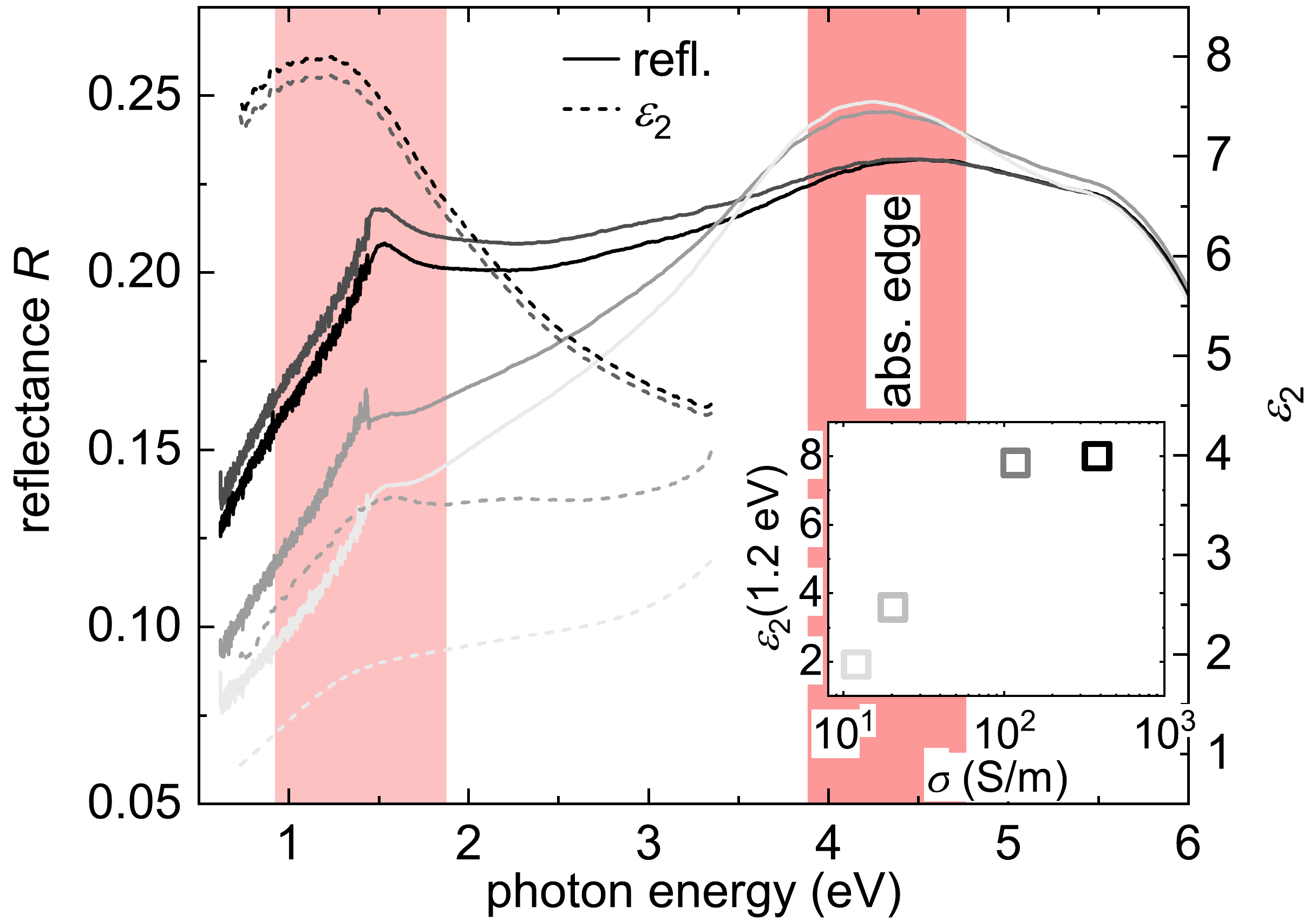}
	\caption{Measured optical reflectance $R$ and imaginary part $\varepsilon_2$ of the dielectric function obtained by ellipsometric modeling for a series of reactively sputtered NiO thin films on fused silica substrates grown at different oxygen partial pressures. Absorption by defect-induced band and at absorption edge highlighted in red. Inset: relation between defect absorption strengths and the electrical conductivity of the films.}
	\label{fig:NiO_sput_pO2_R+eps2}
\end{figure}

Fig.~\ref{fig:NiO_sput_pO2_R+eps2} shows optical reflectance $R$ and $\varepsilon_2$ spectra for four films grown by reactive magnetron sputtering at different oxygen partial pressures, which lead to varying V\lwr{Ni} concentrations. To be able to interpret $R$ spectra without having to take into account interference fringes, the films were made very thin (between \SIlist[list-final-separator=and,list-units=single]{10;12}{\nano\meter}). Here, one intense absorption band at photon energies \num{1.2}-\SI{1.5}{\electronvolt} and a slight shoulder-like feature at \SI{2.5}{\electronvolt} are seen to form, exhibiting increasing transition strength as the acceptor concentration increases. These features are visible both in the reflectance and the $\varepsilon_2$ data, which is to be expected as both quantities are directly related to the extinction coefficient, i.e. optical absorption. The inset of Fig.~\ref{fig:NiO_sput_pO2_R+eps2} displays the relation between the electrical conductivity of the films and the value of $\varepsilon_2$ at \SI{1.2}{\electronvolt} (as a measure of the magnitude of absorption), demonstrating that the increased optical absorption is correlated with higher $[V\mlwr{Ni}]$. Moreover, with increasing V\lwr{Ni} concentration, the strength of the band-to-band transition as visible in the reflectance data at $\approx\SI{4}{\electronvolt}$ (the fundamental band gap) decreases, in accordance with the redistribution of spectra weight. The higher-order transition at approximately \SI{5.5}{\electronvolt} is unaffected. 
Similar observations have been made by Egbo \textit{et al.} \cite{Egbo2020a} by spectroscopic ellipsometry on a V\lwr{Ni}-doped NiO thin film. A broad absorption band appears with a maximum at around \SI{1.5}{\electronvolt}, extending between \SI{1}{\electronvolt} and \SI{2.7}{\electronvolt}, accompanied by a reduction of the amplitude of $\varepsilon_2$ at the fundamental absorption edge at approximately \SI{4}{\electronvolt}. We suggest to interpret these observations within the same framework that has already been applied to the case of Li-doped NiO. The introduction of Ni vacancies creates a defect level in the band gap which can be populated by optical excitation of electrons from the valence band. With increasing doping, the concentration of this level becomes so high that the electronic transitions into it become similarly strong as the band-to-band-transitions of the undoped material. At the same time, the redistribution of the spectral weight from the conduction band into the defect band lowers the magnitude of $\varepsilon_2$ (and therefore also the reflectance) at the band-to-band transition energy of \SI{4}{\electronvolt}. The higher-order transition is likely due to transitions into the O $3p$ band \cite{Kuiper1989} which is not affected by the Ni deficiency to the same degree as the Ni $3d$ CBM. Egbo \textit{et al.} have also studied the impact of other dopants in NiO (Cu, Ag, besides Li and V\lwr{Ni}) on the optical and electrical properties \cite{Egbo2020} and found the appearance of broad absorption bands in the range 1-\SI{2.5}{\electronvolt} in all cases, with small but noticeable difference between the dopant species.

Although the emergence of these in-gap states upon doping is well-documented, a direct interpretation within a defect picture that can be applied coherently to various dopants has not been suggested until now. Ono \textit{et al.} \cite{Ono2018} believe the absorption in the visible range to be 
from the above-mentioned internal $d$-$d$ transitions of the Ni\upr{2+} ion that are forbidden in the perfect crystal but become allowed when defects distort the lattice. This seems questionable for several reasons: (i) the absorption bands are significantly lower in energy than the originally two strongest $d$-$d$ lines at \SI{1.75}{\electronvolt} and \SI{3.25}{\electronvolt}, (ii) the significant broadening of the acceptor-induced features compared to the Ni\upr{2+} lines points toward a band-related origin rather than a local phenomenon like intra-ionic excitation, (iii) the doping-induced features develop independently of those associated with Ni\upr{2+} transitions in the pure material, as demonstrated in Ref.~\cite{Allen1976}, and (iv) the observed difference of absorption band energies between dopant species cannot be explained within this framework. Ghosh \textit{et al.} have attributed optical absorption starting from \SI{0.8}{\electronvolt} to electronic transitions from the VB into a low-lying Ni $4s$ CB. However, the fact that this absorption gradually vanishes at elevated temperatures of a few hundred \si{\kelvin} remains unexplainable within this model. Intriguingly, it agrees well with previous studies on the thermal instability of high concentrations of intrinsic defects (namely V\lwr{Ni}) in NiO \cite{Karsthof2020a,Gutierrez2020}: elevated temperatures mobilize the V\lwr{Ni} defects that ultimately reach the film surface where they disappear. This supports the attribution of the absorption band directly to the defect population.

\subsection{Light emission}

\begin{figure}
	\centering
	\includegraphics[width=\columnwidth]{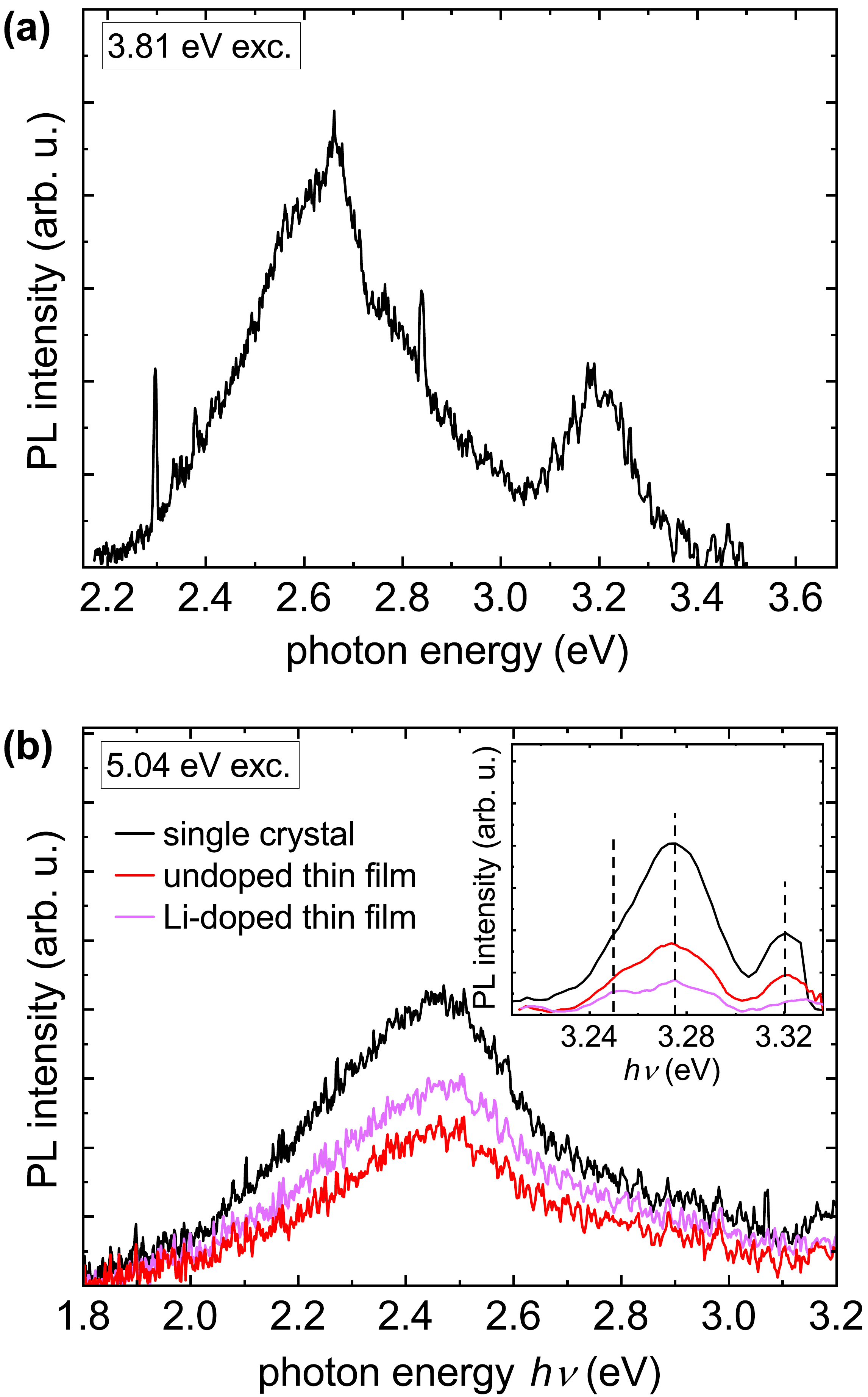}
	\caption{Photoluminescence spectra recorded at \SI{10}{\kelvin} of a NiO single crystal, an undoped and a Li-doped epitaxial layer on MgO substrate. (a) sub-band gap and (b) above-band gap excitation. Inset of (b): enlarged view of intense narrow emission lines likely originating from bound excitons.}
	\label{fig:PL_both_exc}
\end{figure}

Fig.~\ref{fig:PL_both_exc}(a) displays a PL spectrum collected on the (100) surface of a NiO single crystal at \SI{10}{\kelvin} using a \SI{325}{\nano\meter} (\SI{3.81}{\electronvolt}) laser for excitation.  
In the visible spectral range, the PL intensity is generally low, and characterized by a broad emission band in the spectral range \num{2.0}-\SI{3.3}{\electronvolt}, typically peaking at \SIrange[range-units=single]{2.2}{2.4}{\electronvolt}. Under \SI{5.03}{\electronvolt} excitation (Fig~\ref{fig:PL_both_exc}(b)), a narrow and intense multi-component feature appears between \SI{3.24}{\electronvolt} and \SI{3.33}{\electronvolt}. The emission properties of NiO have been studied previously in a few publications \cite{DiazGuerra1997,Volkov2001,Sokolov2012,Ho2015}, reporting on the same spectral features. Sokolov \textit{et al.} \cite{Sokolov2012} have constructed a comprehensive scheme which explains the absorption and emission properties of the perfect NiO lattice by identifying three types of electronic transitions: (i) $p$-$d$ charge transfer (CT), (ii) $d$-$d$ (Ni\upr{2+}) crystal field, and (iii) $d$-$d$ charge transfer transitions. While type (i) and (iii) only occur when exciting above the ``charge transfer gap'' of around \SI{4}{\electronvolt}, the internal Ni\upr{2+} transitions (ii) occur preferably under sub-band gap excitation, such as demonstrated in Ref.~\cite{Volkov2001}. Sokolov \textit{et al.} include in their model the possibility of a relaxation of the $p$-$d$ CT excitons into the localized excited states of type (ii), thereby giving an explanation of the occurrence of emission in the visible spectral range when exciting with above-band gap light. It is, however, interesting to note that when the authors use extreme-UV light ($E\mlwr{exc} = \SI{130}{\electronvolt}$) for excitation, which predominantly excites the bulk material and suppresses emission from defect-rich surfaces, these emission bands disappear completely. A correlation with defects therefore appears likely. The Sokolov model does not explicitly take radiative recombination through electronic defect states into account.

According to Ref.~\cite{Sokolov2012}, the $d$-$d$ CT excitons (iii) recombine radiatively from a narrow-spaced state doublet into the ground state, emitting at energies close to \SI{3.3}{\electronvolt} (I\lwr{1} and I\lwr{2} lines). A more recent work by Ho \textit{et al.} \cite{Ho2015} studied the emission properties NiO nanostructures of high crystalline quality under \SI{4.66}{\electronvolt} excitation for temperatures between \SI{10}{\kelvin} and room temperature. It is claimed that the lines close to \SI{3.3}{\electronvolt} can be resolved into a rich substructure which originates from different bound exciton complexes as well as donor-acceptor pair transitions. The \num{2.2}-\SI{3.0}{\electronvolt} band is entirely absent in their study. Taking into account the high crystalline quality of the samples studied by Ho \textit{et al.}, a correlation of the occurrence of this band with defects appears convincing.

In Fig.~\ref{fig:PL_both_exc}(b), both the intensity of the green band and of the near-band edge emission (NBE) are shown to vary between the single crystal, the undoped and the Li-doped thin films. While the absolute measured intensity can depend on various sample properties such as surface roughness or absorption coefficient, the ratio between the (defect-related) green band and the NBE is mainly affected by defect concentrations. Lower defect densities lead to longer exciton lifetimes because of decreased non-radiative recombination rates, which in turn increases NBE luminescence. Therefore, we suggest to assign the \SIrange[range-units=single]{2.2}{3.0}{\electronvolt} band to radiative carrier recombination via defect states, while the \SI{3.3}{\electronvolt} multiplet is caused by the recombination of excitons bound to defects.

\subsection{Hybrid-functional calculations}

\begin{figure}
	\centering
	\includegraphics[width=0.85\columnwidth]{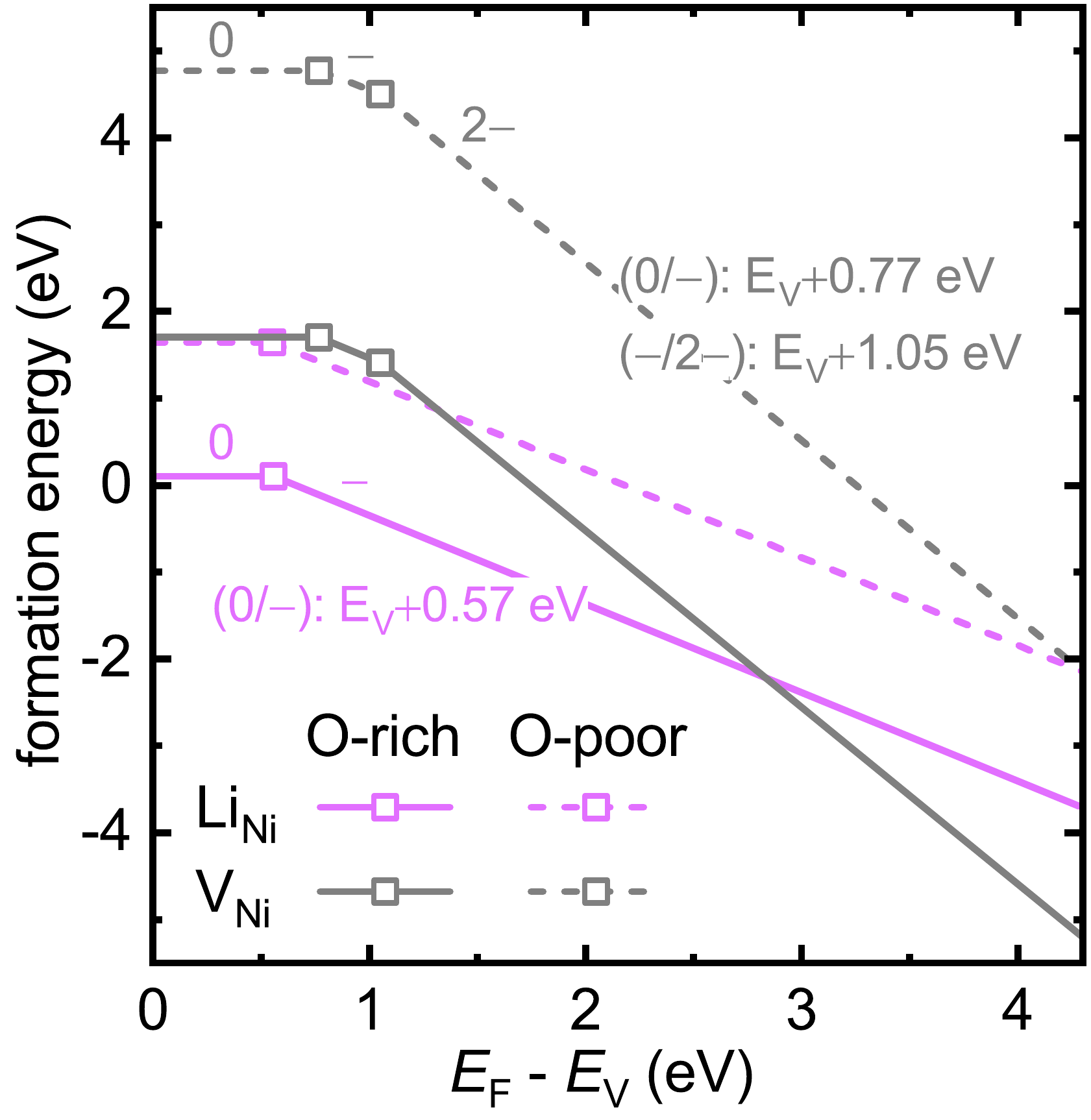}
	\caption{Formation energy of Li\lwr{Ni} and V\lwr{Ni} in NiO under O-rich (solid lines) and Ni-rich (dashed lines) conditions, as a function of Fermi-level position.}
	\label{fig:formation_energies}
\end{figure}

Fig.~\ref{fig:formation_energies} shows the formation energy of Li\lwr{Ni} and V\lwr{Ni}, under O- and Ni-rich conditions, as a function of the Fermi level position in the band gap. Li\lwr{Ni} and V\lwr{Ni} act as single and double acceptors, respectively, and exhibit very low formation energies. This is consistent with the high solubility of Li in NiO and the observed $p$-type conductivity in samples grown under O-rich conditions or doped with Li. Previous calculations based on hybrid and semilocal functionals also indicate
that the concentration of the potential hole killer V\lwr{O} will be low due to a high formation energy \cite{Lany2007,Dawson2015}.
The predicted $(0/-)$ levels of Li\lwr{Ni} and V\lwr{Ni} occur at \SI{0.57}{\electronvolt} and \SI{0.77}{\electronvolt}, respectively, and are both approximately \SI{0.36}{\electronvolt} deeper than those obtained experimentally by means of thermal admittance spectroscopy (TAS) measurements on samples grown to be rich in Li\lwr{Ni} and V\lwr{Ni}, respectively (\SI{0.19}{\electronvolt} and \SI{0.41}{\electronvolt}, assigned to the $(0/-)$ levels \cite{Karsthof2020}). It should be noted, however, that the calculated levels represent the dilute limit (isolated defects). At high concentrations, the interaction between defects can lead to the formation of defect bands. In the case of Li, the effects of alloying on the electronic and crystal structure are also not included in our calculations.
The relative positions of the calculated $(0/-)$ levels of Li\lwr{Ni} and V\lwr{Ni} (energy separation of \SI{0.20}{\electronvolt}), however, are in good agreement with the experimental data (\SI{0.22}{\electronvolt} difference \cite{Karsthof2020}). 
The present calculations also show that by introducing local lattice distortions around a Ni atom, a hole can become trapped in a localized state, i.e., a small polaron forms. The stability of the localized hole (the so-called hole self-trapping energy, $E\mlwr{ST}$) is given by the total energy difference between a localized hole (including the cost of the local lattice distortion) and a delocalized hole in the perfect lattice \cite{Varley2010,Varley2012}. The calculated $E\mlwr{ST}$ is \SI{0.05}{\electronvolt}.

\begin{figure}
	\centering
	\includegraphics[width=0.85\columnwidth]{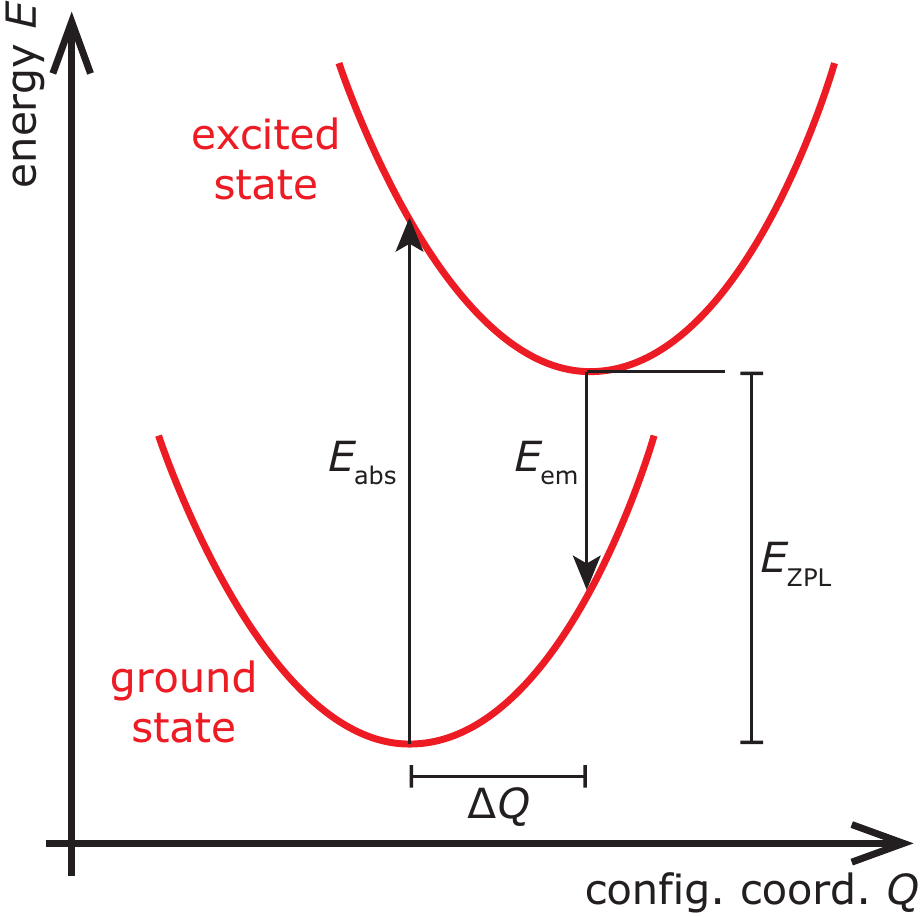}
	\caption{Schematic 1D CC diagram, demonstrating the transitions between ground and excited states of an idealized electronic state, with absorption, emission and zero-phonon line energies $E\mlwr{abs}$, $E\mlwr{em}$, $E\mlwr{ZPL}$ and the CC displacement $\Delta Q$.}
	\label{fig:CC_diagram}
\end{figure}

To investigate the optical transition energies of the Li\lwr{Ni} and V\lwr{Ni} acceptors, 1D configuration coordinate (CC) diagrams as schematically shown in Fig.~\ref{fig:CC_diagram}(a) were constructed. As ground and excited states we take into account different charge states of the Li\lwr{Ni} and V\lwr{Ni} defects, alongside with VB holes or CB electrons where applicable. The resulting parameter values are collected in Table~\ref{tab:CC_diag_values}. We start by considering the optical transition of an electron from the valence band maximum to the empty polaronic hole states of Li\lwr{Ni} and V\lwr{Ni}. For Li\lwr{Ni}, which has only one acceptor level, the only relevant transition is $\text{Li}\muplwr{0}{Ni}+ h\nu \rightarrow \text{Li}\muplwr{-}{Ni} + h\muplwr{+}{VBM}$. This results in an absorption energy ($E\mlwr{abs}$) of \SI{1.48}{\electronvolt}. For V\lwr{Ni}, there are two transitions to consider: (i) $\text{V}\muplwr{0}{Ni}+ h\nu \rightarrow \text{V}\muplwr{-}{Ni} + h\muplwr{+}{VBM}$ and (ii) $\text{V}\muplwr{-}{Ni}+ h\nu \rightarrow \text{V}\muplwr{2-}{Ni} + h\muplwr{+}{VBM}$. Here, the first transition is associated with an absorption energy of \SI{1.51}{\electronvolt}, which is close to the value obtained for the corresponding transition of Li\lwr{Ni}. The second transition yields a higher absorption energy of \SI{1.88}{\electronvolt}.

\begin{table}
	\caption{Parameter values determined from the configuration coordinate diagrams for different electronic states associated with defects: absorption, emission and zero-phonon line energies $E\mlwr{abs}$, $E\mlwr{em}$, $E\mlwr{ZPL}$ (all in \si{\electronvolt}), and change in configuration coordinate $\Delta Q$ (in \si{\amu\tothe{1/2}\angstrom}). Important absorption and emission energies to be compared with experiment are highlighted in bold-face font.}
	\label{tab:CC_diag_values}
	\sisetup{detect-weight,
		         mode=text, 
		         table-format=-2.2
		         }
	\begin{tabular}{ccSSSS}
	\toprule
	GS & ES & {$E\mlwr{abs}$} & {$E\mlwr{em}$} & {$E\mlwr{ZPL}$} & {$\Delta Q$} \\
	\midrule
	Li\uplwr{0}{Ni} & Li\uplwr{-}{Ni} + $h$\uplwr{+}{VBM} & \B 1.48 & -0.31 & 0.57 & 1.55  \\
	V\uplwr{0}{Ni} & V\uplwr{-}{Ni} + $h$\uplwr{+}{VBM} & \B 1.51 & -0.05 & 0.77 & 1.39 \\
	V\uplwr{-}{Ni} & V\uplwr{2-}{Ni} + $h$\uplwr{+}{VBM} & \B 1.88 & 0.35 & 1.05 & 1.42 \\ 
	\midrule
	Li\uplwr{-}{Ni} & Li\uplwr{0}{Ni} + $e$\uplwr{-}{CBM} & 4.61 & \B 2.81 & 3.73 & 1.55 \\
	V\uplwr{-}{Ni} & V\uplwr{0}{Ni} + $e$\uplwr{-}{CBM} & 4.35 & \B 2.79 & 3.53 & 1.39  \\
	V\uplwr{2-}{Ni} & V\uplwr{-}{Ni} + $e$\uplwr{-}{CBM} & 3.94 & \B 2.42 & 3.24 & 1.42 \\
	\midrule
	bulk & SP + $e$\uplwr{-}{CBM} & 5.22 & \B 3.28 & 4.25 & 1.62 \\
	\bottomrule
	\end{tabular}
\end{table}

The strong polaronic distortion ($\Delta Q \approx$ \SIrange[range-phrase=-,range-units=single]{1.3}{1.6}{\amu\tothe{1/2}\angstrom}) results in large relaxation energies (Franck-Condon shifts) in the \SIrange[range-phrase=-,range-units=single]{0.6}{1}{\electronvolt} range (depending on the defect). Such strong electron-phonon coupling can be expected to result in significant vibrational broadening of the absorption onsets, in line with experimental data \cite{Zhang2018,Frodason2019}. In this case, the calculated absorption energies of the order of \SI{1.48}{\electronvolt} and \SIrange[range-phrase=-,range-units=single]{1.5}{1.8}{\electronvolt} for Li\lwr{Ni} and V\lwr{Ni}, respectively, reproduces the experimental observations well.
Note that the present calculations only consider transitions from the VBM to the polaronic hole states, i.e., transitions of electrons from lower-lying VB states are not included \cite{Zhang2018}.

The radiative recombination of an electron in the CBM with a free or acceptor-bound small hole polaron, which reflects the situation in a photoluminescence experiment with above-band gap light, leads to emission of a photon of energy $E\mlwr{em}$. In order to excite acceptors to a state with a free electron in the CB, photon energies of between \SI{3.94}{\electronvolt} and \SI{4.61}{\electronvolt} are predicted to be required. Emission energies are calculated to be in the range \SIrange[range-phrase=-,range-units=single]{2}{3.5}{\electronvolt}. The $(0/-)$ transitions of Li\lwr{Ni} and V\lwr{Ni} yield very close emission energies of \SI{2.81}{\electronvolt} and \SI{2.79}{\electronvolt}, respectively, while the $(-/2-)$ transition of V\lwr{Ni} results in a lower emission energy of \SI{2.42}{\electronvolt}. Considering the low formation energy of V\lwr{Ni}, this defect is a good candidate for the observed visible luminescence at the low energy side of the broad emission.
For the small hole polaron (or self-trapped hole), the calculated emission energy is \SI{3.28}{\electronvolt}, which is consistent with the \SI{3.2}{\electronvolt} band observed here and in the literature and its assignment by Sokolov \textit{et al.} \cite{Sokolov2012}. For the process of generating the small polaron-free electron pair, above-band gap excitation energies are necessary, explaining why the \SI{3.2}{\electronvolt} multiplet appeared much more strongly when exciting above \SI{5}{\electronvolt}.

\section{Conclusion}

By comparing experimental results with DFT calculations, it could be shown that the optical absorption and light emission properties of NiO are dominated by defects present in the sample. Specifically, this work quantitatively describes the strong anti-correlation of electrical conductivity and optical transmittance of doped NiO, which has long been known but lacked an explanation of its origin. Even though the main acceptors, such as Li\lwr{Ni} and V\lwr{Ni}, possess fairly shallow electronic states that can be at least partially ionized at room temperature, the strong electron-phonon interaction leads to a significant blueshift of the optical transitions between different defect charge states. This renders heavily doped NiO essentially a narrow-gap semiconductor, with band gap values of slightly above \SI{1}{\electronvolt}. The results of the calculations are additionally tested by comparing the predicted light emission to experimental data, and excellent agreement has been found.

%

\begin{acknowledgments}
	This work was funded by S\"{a}chsische Aufbaubank (SAB, project No. 100112104), Deutsche Forschungsgemeinschaft (DFG) in the framework of the collaborative research center ``SFB 762: Functionality of Oxidic Interfaces'' (project No. 31047526), and the Research Council of Norway (FUNDAMeNT, project no. 251131, NoSiCaR, project no. 274742, GoPOW, project no. 314017, FME SUSOLTECH, project no. 257639). The computations were performed on resources provided by UNINETT Sigma2, the National Infrastructure for High Performance Computing and Data Storage in Norway. The Research Council of Norway is acknowledged for the support to the Norwegian Micro- and Nano-Fabrication Facility, NorFab (project no. 295864).
\end{acknowledgments}

\section*{Conflicts of interest}

The authors have no conflict of interest to declare.

\section*{References}
%



\begin{thebibliography}{42}%
\makeatletter
\providecommand \@ifxundefined [1]{%
 \@ifx{#1\undefined}
}%
\providecommand \@ifnum [1]{%
 \ifnum #1\expandafter \@firstoftwo
 \else \expandafter \@secondoftwo
 \fi
}%
\providecommand \@ifx [1]{%
 \ifx #1\expandafter \@firstoftwo
 \else \expandafter \@secondoftwo
 \fi
}%
\providecommand \natexlab [1]{#1}%
\providecommand \enquote  [1]{``#1''}%
\providecommand \bibnamefont  [1]{#1}%
\providecommand \bibfnamefont [1]{#1}%
\providecommand \citenamefont [1]{#1}%
\providecommand \href@noop [0]{\@secondoftwo}%
\providecommand \href [0]{\begingroup \@sanitize@url \@href}%
\providecommand \@href[1]{\@@startlink{#1}\@@href}%
\providecommand \@@href[1]{\endgroup#1\@@endlink}%
\providecommand \@sanitize@url [0]{\catcode `\\12\catcode `\$12\catcode
  `\&12\catcode `\#12\catcode `\^12\catcode `\_12\catcode `\%12\relax}%
\providecommand \@@startlink[1]{}%
\providecommand \@@endlink[0]{}%
\providecommand \url  [0]{\begingroup\@sanitize@url \@url }%
\providecommand \@url [1]{\endgroup\@href {#1}{\urlprefix }}%
\providecommand \urlprefix  [0]{URL }%
\providecommand \Eprint [0]{\href }%
\providecommand \doibase [0]{https://doi.org/}%
\providecommand \selectlanguage [0]{\@gobble}%
\providecommand \bibinfo  [0]{\@secondoftwo}%
\providecommand \bibfield  [0]{\@secondoftwo}%
\providecommand \translation [1]{[#1]}%
\providecommand \BibitemOpen [0]{}%
\providecommand \bibitemStop [0]{}%
\providecommand \bibitemNoStop [0]{.\EOS\space}%
\providecommand \EOS [0]{\spacefactor3000\relax}%
\providecommand \BibitemShut  [1]{\csname bibitem#1\endcsname}%
\let\auto@bib@innerbib\@empty
\bibitem [{\citenamefont {Xu}\ \emph {et~al.}(2019)\citenamefont {Xu},
  \citenamefont {Chen}, \citenamefont {Jin}, \citenamefont {Liu}, \citenamefont
  {Dong}, \citenamefont {Bai}, \citenamefont {Song},\ and\ \citenamefont
  {Reiss}}]{Xu2019}%
  \BibitemOpen
  \bibfield  {author} {\bibinfo {author} {\bibfnamefont {L.}~\bibnamefont
  {Xu}}, \bibinfo {author} {\bibfnamefont {X.}~\bibnamefont {Chen}}, \bibinfo
  {author} {\bibfnamefont {J.}~\bibnamefont {Jin}}, \bibinfo {author}
  {\bibfnamefont {W.}~\bibnamefont {Liu}}, \bibinfo {author} {\bibfnamefont
  {B.}~\bibnamefont {Dong}}, \bibinfo {author} {\bibfnamefont {X.}~\bibnamefont
  {Bai}}, \bibinfo {author} {\bibfnamefont {H.}~\bibnamefont {Song}},\ and\
  \bibinfo {author} {\bibfnamefont {P.}~\bibnamefont {Reiss}},\ }\bibfield
  {title} {\enquote {\bibinfo {title} {Inverted perovskite solar cells
  employing doped {NiO} hole transport layers: A review},}\ }\href
  {https://doi.org/10.1016/j.nanoen.2019.103860} {\bibfield  {journal}
  {\bibinfo  {journal} {Nano Energy}\ }\textbf {\bibinfo {volume} {63}},\
  \bibinfo {pages} {103860} (\bibinfo {year} {2019})}\BibitemShut {NoStop}%
\bibitem [{\citenamefont {Li}\ \emph {et~al.}(2019)\citenamefont {Li},
  \citenamefont {Wang}, \citenamefont {Chen}, \citenamefont {Cui},
  \citenamefont {Ding}, \citenamefont {Li}, \citenamefont {Zhang},
  \citenamefont {Zhao},\ and\ \citenamefont {Zhang}}]{Li2019}%
  \BibitemOpen
  \bibfield  {author} {\bibinfo {author} {\bibfnamefont {R.}~\bibnamefont
  {Li}}, \bibinfo {author} {\bibfnamefont {P.}~\bibnamefont {Wang}}, \bibinfo
  {author} {\bibfnamefont {B.}~\bibnamefont {Chen}}, \bibinfo {author}
  {\bibfnamefont {X.}~\bibnamefont {Cui}}, \bibinfo {author} {\bibfnamefont
  {Y.}~\bibnamefont {Ding}}, \bibinfo {author} {\bibfnamefont {Y.}~\bibnamefont
  {Li}}, \bibinfo {author} {\bibfnamefont {D.}~\bibnamefont {Zhang}}, \bibinfo
  {author} {\bibfnamefont {Y.}~\bibnamefont {Zhao}},\ and\ \bibinfo {author}
  {\bibfnamefont {X.}~\bibnamefont {Zhang}},\ }\bibfield  {title} {\enquote
  {\bibinfo {title} {{NiO}$_x$/spiro hole transport bilayers for stable
  perovskite solar cells with efficiency exceeding 21{\%}},}\ }\href
  {https://doi.org/10.1021/acsenergylett.9b02112} {\bibfield  {journal}
  {\bibinfo  {journal} {{ACS} Energy Letters}\ }\textbf {\bibinfo {volume}
  {5}},\ \bibinfo {pages} {79--86} (\bibinfo {year} {2019})}\BibitemShut
  {NoStop}%
\bibitem [{\citenamefont {Feng}\ \emph {et~al.}(2020)\citenamefont {Feng},
  \citenamefont {Wang}, \citenamefont {Zhou}, \citenamefont {Li}, \citenamefont
  {Wang}, \citenamefont {Zang},\ and\ \citenamefont {Chen}}]{Feng2020}%
  \BibitemOpen
  \bibfield  {author} {\bibinfo {author} {\bibfnamefont {M.}~\bibnamefont
  {Feng}}, \bibinfo {author} {\bibfnamefont {M.}~\bibnamefont {Wang}}, \bibinfo
  {author} {\bibfnamefont {H.}~\bibnamefont {Zhou}}, \bibinfo {author}
  {\bibfnamefont {W.}~\bibnamefont {Li}}, \bibinfo {author} {\bibfnamefont
  {S.}~\bibnamefont {Wang}}, \bibinfo {author} {\bibfnamefont {Z.}~\bibnamefont
  {Zang}},\ and\ \bibinfo {author} {\bibfnamefont {S.}~\bibnamefont {Chen}},\
  }\bibfield  {title} {\enquote {\bibinfo {title} {High-efficiency and stable
  inverted planar perovskite solar cells with pulsed laser deposited cu-doped
  {NiO}$_x$ hole-transport layers},}\ }\href
  {https://doi.org/10.1021/acsami.0c15923} {\bibfield  {journal} {\bibinfo
  {journal} {{ACS} Applied Materials {\&} Interfaces}\ }\textbf {\bibinfo
  {volume} {12}},\ \bibinfo {pages} {50684--50691} (\bibinfo {year}
  {2020})}\BibitemShut {NoStop}%
\bibitem [{\citenamefont {Biela{\'{n}}ski}\ \emph {et~al.}(1962)\citenamefont
  {Biela{\'{n}}ski}, \citenamefont {Dere{\'{n}}}, \citenamefont {Haber},\ and\
  \citenamefont {S{\l}oczy{\'{n}}ski}}]{Bielanski1962}%
  \BibitemOpen
  \bibfield  {author} {\bibinfo {author} {\bibfnamefont {A.}~\bibnamefont
  {Biela{\'{n}}ski}}, \bibinfo {author} {\bibfnamefont {J.}~\bibnamefont
  {Dere{\'{n}}}}, \bibinfo {author} {\bibfnamefont {J.}~\bibnamefont {Haber}},\
  and\ \bibinfo {author} {\bibfnamefont {J.}~\bibnamefont
  {S{\l}oczy{\'{n}}ski}},\ }\bibfield  {title} {\enquote {\bibinfo {title}
  {Physico-chemical properties of alkali- and iron-doped nickel oxide},}\
  }\href {https://doi.org/10.1039/tf9625800166} {\ \textbf {\bibinfo {volume}
  {58}},\ \bibinfo {pages} {166--175} (\bibinfo {year} {1962})}\BibitemShut
  {NoStop}%
\bibitem [{\citenamefont {Lany}, \citenamefont {Osorio-Guill{\'{e}}n},\ and\
  \citenamefont {Zunger}(2007)}]{Lany2007}%
  \BibitemOpen
  \bibfield  {author} {\bibinfo {author} {\bibfnamefont {S.}~\bibnamefont
  {Lany}}, \bibinfo {author} {\bibfnamefont {J.}~\bibnamefont
  {Osorio-Guill{\'{e}}n}},\ and\ \bibinfo {author} {\bibfnamefont
  {A.}~\bibnamefont {Zunger}},\ }\bibfield  {title} {\enquote {\bibinfo {title}
  {Origins of the doping asymmetry in oxides: Hole doping in {NiO} versus
  electron doping in {ZnO}},}\ }\href
  {https://doi.org/10.1103/physrevb.75.241203} {\bibfield  {journal} {\bibinfo
  {journal} {Physical Review B}\ }\textbf {\bibinfo {volume} {75}},\ \bibinfo
  {pages} {241203(R)} (\bibinfo {year} {2007})}\BibitemShut {NoStop}%
\bibitem [{\citenamefont {Yang}\ \emph {et~al.}(2012)\citenamefont {Yang},
  \citenamefont {Pu}, \citenamefont {Zhou},\ and\ \citenamefont
  {Zhang}}]{Yang2012}%
  \BibitemOpen
  \bibfield  {author} {\bibinfo {author} {\bibfnamefont {M.}~\bibnamefont
  {Yang}}, \bibinfo {author} {\bibfnamefont {H.}~\bibnamefont {Pu}}, \bibinfo
  {author} {\bibfnamefont {Q.}~\bibnamefont {Zhou}},\ and\ \bibinfo {author}
  {\bibfnamefont {Q.}~\bibnamefont {Zhang}},\ }\bibfield  {title} {\enquote
  {\bibinfo {title} {Transparent p-type conducting {K}-doped {NiO} films
  deposited by pulsed plasma deposition},}\ }\href
  {https://doi.org/10.1016/j.tsf.2012.05.005} {\bibfield  {journal} {\bibinfo
  {journal} {Thin Solid Films}\ }\textbf {\bibinfo {volume} {520}},\ \bibinfo
  {pages} {5884--5888} (\bibinfo {year} {2012})}\BibitemShut {NoStop}%
\bibitem [{\citenamefont {Zhang}\ \emph {et~al.}(2018)\citenamefont {Zhang},
  \citenamefont {Li}, \citenamefont {Hoye}, \citenamefont {MacManus-Driscoll},
  \citenamefont {Budde}, \citenamefont {Bierwagen}, \citenamefont {Wang},
  \citenamefont {Du}, \citenamefont {Wahila}, \citenamefont {Piper},
  \citenamefont {Lee}, \citenamefont {Edwards}, \citenamefont {Dhanak},\ and\
  \citenamefont {Zhang}}]{Zhang2018}%
  \BibitemOpen
  \bibfield  {author} {\bibinfo {author} {\bibfnamefont {J.~Y.}\ \bibnamefont
  {Zhang}}, \bibinfo {author} {\bibfnamefont {W.~W.}\ \bibnamefont {Li}},
  \bibinfo {author} {\bibfnamefont {R.~L.~Z.}\ \bibnamefont {Hoye}}, \bibinfo
  {author} {\bibfnamefont {J.~L.}\ \bibnamefont {MacManus-Driscoll}}, \bibinfo
  {author} {\bibfnamefont {M.}~\bibnamefont {Budde}}, \bibinfo {author}
  {\bibfnamefont {O.}~\bibnamefont {Bierwagen}}, \bibinfo {author}
  {\bibfnamefont {L.}~\bibnamefont {Wang}}, \bibinfo {author} {\bibfnamefont
  {Y.}~\bibnamefont {Du}}, \bibinfo {author} {\bibfnamefont {M.~J.}\
  \bibnamefont {Wahila}}, \bibinfo {author} {\bibfnamefont {L.~F.~J.}\
  \bibnamefont {Piper}}, \bibinfo {author} {\bibfnamefont {T.-L.}\ \bibnamefont
  {Lee}}, \bibinfo {author} {\bibfnamefont {H.~J.}\ \bibnamefont {Edwards}},
  \bibinfo {author} {\bibfnamefont {V.~R.}\ \bibnamefont {Dhanak}},\ and\
  \bibinfo {author} {\bibfnamefont {K.~H.~L.}\ \bibnamefont {Zhang}},\
  }\bibfield  {title} {\enquote {\bibinfo {title} {Electronic and transport
  properties of {Li}-doped {NiO} epitaxial thin films},}\ }\href
  {https://doi.org/10.1039/c7tc05331b} {\bibfield  {journal} {\bibinfo
  {journal} {J. Mat. Chem. C}\ }\textbf {\bibinfo {volume} {6}},\ \bibinfo
  {pages} {2275--2282} (\bibinfo {year} {2018})}\BibitemShut {NoStop}%
\bibitem [{\citenamefont {Egbo}\ \emph
  {et~al.}(2020{\natexlab{a}})\citenamefont {Egbo}, \citenamefont {Ekuma},
  \citenamefont {Liu},\ and\ \citenamefont {Yu}}]{Egbo2020}%
  \BibitemOpen
  \bibfield  {author} {\bibinfo {author} {\bibfnamefont {K.~O.}\ \bibnamefont
  {Egbo}}, \bibinfo {author} {\bibfnamefont {C.~E.}\ \bibnamefont {Ekuma}},
  \bibinfo {author} {\bibfnamefont {C.~P.}\ \bibnamefont {Liu}},\ and\ \bibinfo
  {author} {\bibfnamefont {K.~M.}\ \bibnamefont {Yu}},\ }\bibfield  {title}
  {\enquote {\bibinfo {title} {Efficient p-type doping of sputter-deposited
  {NiO} thin films with {Li}, {Ag}, and {Cu} acceptors},}\ }\href
  {https://doi.org/10.1103/physrevmaterials.4.104603} {\bibfield  {journal}
  {\bibinfo  {journal} {Phys Rev Materials}\ }\textbf {\bibinfo {volume} {4}}
  (\bibinfo {year} {2020}{\natexlab{a}}),\
  10.1103/physrevmaterials.4.104603}\BibitemShut {NoStop}%
\bibitem [{\citenamefont {Egbo}\ \emph
  {et~al.}(2020{\natexlab{b}})\citenamefont {Egbo}, \citenamefont {Liu},
  \citenamefont {Ekuma},\ and\ \citenamefont {Yu}}]{Egbo2020a}%
  \BibitemOpen
  \bibfield  {author} {\bibinfo {author} {\bibfnamefont {K.~O.}\ \bibnamefont
  {Egbo}}, \bibinfo {author} {\bibfnamefont {C.~P.}\ \bibnamefont {Liu}},
  \bibinfo {author} {\bibfnamefont {C.~E.}\ \bibnamefont {Ekuma}},\ and\
  \bibinfo {author} {\bibfnamefont {K.~M.}\ \bibnamefont {Yu}},\ }\bibfield
  {title} {\enquote {\bibinfo {title} {Vacancy defects induced changes in the
  electronic and optical properties of {NiO} studied by spectroscopic
  ellipsometry and first-principles calculations},}\ }\href
  {https://doi.org/10.1063/5.0021650} {\bibfield  {journal} {\bibinfo
  {journal} {J Appl Phys}\ }\textbf {\bibinfo {volume} {128}},\ \bibinfo
  {pages} {135705} (\bibinfo {year} {2020}{\natexlab{b}})}\BibitemShut
  {NoStop}%
\bibitem [{\citenamefont {Karsthof}\ \emph
  {et~al.}(2020{\natexlab{a}})\citenamefont {Karsthof}, \citenamefont {von
  Wenckstern}, \citenamefont {Olsen},\ and\ \citenamefont
  {Grundmann}}]{Karsthof2020}%
  \BibitemOpen
  \bibfield  {author} {\bibinfo {author} {\bibfnamefont {R.}~\bibnamefont
  {Karsthof}}, \bibinfo {author} {\bibfnamefont {H.}~\bibnamefont {von
  Wenckstern}}, \bibinfo {author} {\bibfnamefont {V.~S.}\ \bibnamefont
  {Olsen}},\ and\ \bibinfo {author} {\bibfnamefont {M.}~\bibnamefont
  {Grundmann}},\ }\bibfield  {title} {\enquote {\bibinfo {title}
  {Identification of {Li$_{\mathrm{Ni}}$} and {V$_{\mathrm{Ni}}$} acceptor
  levels in doped nickel oxide},}\ }\href {https://doi.org/10.1063/5.0032102}
  {\bibfield  {journal} {\bibinfo  {journal} {{APL} Materials}\ }\textbf
  {\bibinfo {volume} {8}},\ \bibinfo {pages} {121106} (\bibinfo {year}
  {2020}{\natexlab{a}})}\BibitemShut {NoStop}%
\bibitem [{\citenamefont {Bl\"ochl}(1994)}]{Bloechl1994}%
  \BibitemOpen
  \bibfield  {author} {\bibinfo {author} {\bibfnamefont {P.~E.}\ \bibnamefont
  {Bl\"ochl}},\ }\bibfield  {title} {\enquote {\bibinfo {title} {Projector
  augmented-wave method},}\ }\href {https://doi.org/10.1103/physrevb.50.17953}
  {\bibfield  {journal} {\bibinfo  {journal} {Physical Review B}\ }\textbf
  {\bibinfo {volume} {50}},\ \bibinfo {pages} {17953--17979} (\bibinfo {year}
  {1994})}\BibitemShut {NoStop}%
\bibitem [{\citenamefont {Kresse}\ and\ \citenamefont
  {Joubert}(1999)}]{Kresse1999}%
  \BibitemOpen
  \bibfield  {author} {\bibinfo {author} {\bibfnamefont {G.}~\bibnamefont
  {Kresse}}\ and\ \bibinfo {author} {\bibfnamefont {D.}~\bibnamefont
  {Joubert}},\ }\bibfield  {title} {\enquote {\bibinfo {title} {From ultrasoft
  pseudopotentials to the projector augmented-wave method},}\ }\href
  {https://doi.org/10.1103/physrevb.59.1758} {\bibfield  {journal} {\bibinfo
  {journal} {Physical Review B}\ }\textbf {\bibinfo {volume} {59}},\ \bibinfo
  {pages} {1758--1775} (\bibinfo {year} {1999})}\BibitemShut {NoStop}%
\bibitem [{\citenamefont {Kresse}\ and\ \citenamefont
  {Furthm\"uller}(1996)}]{Kresse1996}%
  \BibitemOpen
  \bibfield  {author} {\bibinfo {author} {\bibfnamefont {G.}~\bibnamefont
  {Kresse}}\ and\ \bibinfo {author} {\bibfnamefont {J.}~\bibnamefont
  {Furthm\"uller}},\ }\bibfield  {title} {\enquote {\bibinfo {title} {Efficient
  iterative schemes for ab initio total-energy calculations using a plane-wave
  basis set},}\ }\href {https://doi.org/10.1103/physrevb.54.11169} {\bibfield
  {journal} {\bibinfo  {journal} {Physical Review B}\ }\textbf {\bibinfo
  {volume} {54}},\ \bibinfo {pages} {11169--11186} (\bibinfo {year}
  {1996})}\BibitemShut {NoStop}%
\bibitem [{\citenamefont {Liu}\ \emph {et~al.}(2019)\citenamefont {Liu},
  \citenamefont {Franchini}, \citenamefont {Marsman},\ and\ \citenamefont
  {Kresse}}]{Liu2019}%
  \BibitemOpen
  \bibfield  {author} {\bibinfo {author} {\bibfnamefont {P.}~\bibnamefont
  {Liu}}, \bibinfo {author} {\bibfnamefont {C.}~\bibnamefont {Franchini}},
  \bibinfo {author} {\bibfnamefont {M.}~\bibnamefont {Marsman}},\ and\ \bibinfo
  {author} {\bibfnamefont {G.}~\bibnamefont {Kresse}},\ }\bibfield  {title}
  {\enquote {\bibinfo {title} {Assessing model-dielectric-dependent hybrid
  functionals on the antiferromagnetic transition-metal monoxides {MnO}, {FeO},
  {CoO}, and {NiO}},}\ }\href {https://doi.org/10.1088/1361-648x/ab4150}
  {\bibfield  {journal} {\bibinfo  {journal} {Journal of Physics: Condensed
  Matter}\ }\textbf {\bibinfo {volume} {32}},\ \bibinfo {pages} {015502}
  (\bibinfo {year} {2019})}\BibitemShut {NoStop}%
\bibitem [{\citenamefont {Krukau}\ \emph {et~al.}(2006)\citenamefont {Krukau},
  \citenamefont {Vydrov}, \citenamefont {Izmaylov},\ and\ \citenamefont
  {Scuseria}}]{Krukau2006}%
  \BibitemOpen
  \bibfield  {author} {\bibinfo {author} {\bibfnamefont {A.~V.}\ \bibnamefont
  {Krukau}}, \bibinfo {author} {\bibfnamefont {O.~A.}\ \bibnamefont {Vydrov}},
  \bibinfo {author} {\bibfnamefont {A.~F.}\ \bibnamefont {Izmaylov}},\ and\
  \bibinfo {author} {\bibfnamefont {G.~E.}\ \bibnamefont {Scuseria}},\
  }\bibfield  {title} {\enquote {\bibinfo {title} {Influence of the exchange
  screening parameter on the performance of screened hybrid functionals},}\
  }\href {https://doi.org/10.1063/1.2404663} {\bibfield  {journal} {\bibinfo
  {journal} {The Journal of Chemical Physics}\ }\textbf {\bibinfo {volume}
  {125}},\ \bibinfo {pages} {224106} (\bibinfo {year} {2006})}\BibitemShut
  {NoStop}%
\bibitem [{\citenamefont {Freysoldt}\ \emph {et~al.}(2014)\citenamefont
  {Freysoldt}, \citenamefont {Grabowski}, \citenamefont {Hickel}, \citenamefont
  {Neugebauer}, \citenamefont {Kresse}, \citenamefont {Janotti},\ and\
  \citenamefont {de~Walle}}]{Freysoldt2014}%
  \BibitemOpen
  \bibfield  {author} {\bibinfo {author} {\bibfnamefont {C.}~\bibnamefont
  {Freysoldt}}, \bibinfo {author} {\bibfnamefont {B.}~\bibnamefont
  {Grabowski}}, \bibinfo {author} {\bibfnamefont {T.}~\bibnamefont {Hickel}},
  \bibinfo {author} {\bibfnamefont {J.}~\bibnamefont {Neugebauer}}, \bibinfo
  {author} {\bibfnamefont {G.}~\bibnamefont {Kresse}}, \bibinfo {author}
  {\bibfnamefont {A.}~\bibnamefont {Janotti}},\ and\ \bibinfo {author}
  {\bibfnamefont {C.~G.~V.}\ \bibnamefont {de~Walle}},\ }\bibfield  {title}
  {\enquote {\bibinfo {title} {First-principles calculations for point defects
  in solids},}\ }\href {https://doi.org/10.1103/revmodphys.86.253} {\bibfield
  {journal} {\bibinfo  {journal} {Reviews of Modern Physics}\ }\textbf
  {\bibinfo {volume} {86}},\ \bibinfo {pages} {253--305} (\bibinfo {year}
  {2014})}\BibitemShut {NoStop}%
\bibitem [{\citenamefont {Freysoldt}, \citenamefont {Neugebauer},\ and\
  \citenamefont {de~Walle}(2009)}]{Freysoldt2009}%
  \BibitemOpen
  \bibfield  {author} {\bibinfo {author} {\bibfnamefont {C.}~\bibnamefont
  {Freysoldt}}, \bibinfo {author} {\bibfnamefont {J.}~\bibnamefont
  {Neugebauer}},\ and\ \bibinfo {author} {\bibfnamefont {C.~G.~V.}\
  \bibnamefont {de~Walle}},\ }\bibfield  {title} {\enquote {\bibinfo {title}
  {Fully ab initio finite-size corrections for charged-defect supercell
  calculations},}\ }\href {https://doi.org/10.1103/physrevlett.102.016402}
  {\bibfield  {journal} {\bibinfo  {journal} {Physical Review Letters}\
  }\textbf {\bibinfo {volume} {102}},\ \bibinfo {pages} {016402} (\bibinfo
  {year} {2009})}\BibitemShut {NoStop}%
\bibitem [{\citenamefont {Kumagai}\ and\ \citenamefont
  {Oba}(2014)}]{Kumagai2014}%
  \BibitemOpen
  \bibfield  {author} {\bibinfo {author} {\bibfnamefont {Y.}~\bibnamefont
  {Kumagai}}\ and\ \bibinfo {author} {\bibfnamefont {F.}~\bibnamefont {Oba}},\
  }\bibfield  {title} {\enquote {\bibinfo {title} {Electrostatics-based
  finite-size corrections for first-principles point defect calculations},}\
  }\href {https://doi.org/10.1103/physrevb.89.195205} {\bibfield  {journal}
  {\bibinfo  {journal} {Physical Review B}\ }\textbf {\bibinfo {volume} {89}},\
  \bibinfo {pages} {195205} (\bibinfo {year} {2014})}\BibitemShut {NoStop}%
\bibitem [{\citenamefont {Gake}\ \emph {et~al.}(2020)\citenamefont {Gake},
  \citenamefont {Kumagai}, \citenamefont {Freysoldt},\ and\ \citenamefont
  {Oba}}]{Gake2020}%
  \BibitemOpen
  \bibfield  {author} {\bibinfo {author} {\bibfnamefont {T.}~\bibnamefont
  {Gake}}, \bibinfo {author} {\bibfnamefont {Y.}~\bibnamefont {Kumagai}},
  \bibinfo {author} {\bibfnamefont {C.}~\bibnamefont {Freysoldt}},\ and\
  \bibinfo {author} {\bibfnamefont {F.}~\bibnamefont {Oba}},\ }\bibfield
  {title} {\enquote {\bibinfo {title} {Finite-size corrections for
  defect-involving vertical transitions in supercell calculations},}\ }\href
  {https://doi.org/10.1103/physrevb.101.020102} {\bibfield  {journal} {\bibinfo
   {journal} {Physical Review B}\ }\textbf {\bibinfo {volume} {101}},\ \bibinfo
  {pages} {020102(R)} (\bibinfo {year} {2020})}\BibitemShut {NoStop}%
\bibitem [{\citenamefont {Gielisse}\ \emph {et~al.}(1965)\citenamefont
  {Gielisse}, \citenamefont {Plendl}, \citenamefont {Mansur}, \citenamefont
  {Marshall}, \citenamefont {Mitra}, \citenamefont {Mykolajewycz},\ and\
  \citenamefont {Smakula}}]{Gielisse1965}%
  \BibitemOpen
  \bibfield  {author} {\bibinfo {author} {\bibfnamefont {P.~J.}\ \bibnamefont
  {Gielisse}}, \bibinfo {author} {\bibfnamefont {J.~N.}\ \bibnamefont
  {Plendl}}, \bibinfo {author} {\bibfnamefont {L.~C.}\ \bibnamefont {Mansur}},
  \bibinfo {author} {\bibfnamefont {R.}~\bibnamefont {Marshall}}, \bibinfo
  {author} {\bibfnamefont {S.~S.}\ \bibnamefont {Mitra}}, \bibinfo {author}
  {\bibfnamefont {R.}~\bibnamefont {Mykolajewycz}},\ and\ \bibinfo {author}
  {\bibfnamefont {A.}~\bibnamefont {Smakula}},\ }\bibfield  {title} {\enquote
  {\bibinfo {title} {Infrared properties of {NiO} and {CoO} and their mixed
  crystals},}\ }\href {https://doi.org/10.1063/1.1714508} {\bibfield  {journal}
  {\bibinfo  {journal} {Journal of Applied Physics}\ }\textbf {\bibinfo
  {volume} {36}},\ \bibinfo {pages} {2446--2450} (\bibinfo {year}
  {1965})}\BibitemShut {NoStop}%
\bibitem [{\citenamefont {Alkauskas}\ \emph {et~al.}(2012)\citenamefont
  {Alkauskas}, \citenamefont {Lyons}, \citenamefont {Steiauf},\ and\
  \citenamefont {de~Walle}}]{Alkauskas2012}%
  \BibitemOpen
  \bibfield  {author} {\bibinfo {author} {\bibfnamefont {A.}~\bibnamefont
  {Alkauskas}}, \bibinfo {author} {\bibfnamefont {J.~L.}\ \bibnamefont
  {Lyons}}, \bibinfo {author} {\bibfnamefont {D.}~\bibnamefont {Steiauf}},\
  and\ \bibinfo {author} {\bibfnamefont {C.~G.~V.}\ \bibnamefont {de~Walle}},\
  }\bibfield  {title} {\enquote {\bibinfo {title} {First-principles
  calculations of luminescence spectrum line shapes for defects in
  semiconductors: The example of {GaN} and {ZnO}},}\ }\href
  {https://doi.org/10.1103/physrevlett.109.267401} {\bibfield  {journal}
  {\bibinfo  {journal} {Physical Review Letters}\ }\textbf {\bibinfo {volume}
  {109}},\ \bibinfo {pages} {267401} (\bibinfo {year} {2012})}\BibitemShut
  {NoStop}%
\bibitem [{\citenamefont {Alkauskas}, \citenamefont {McCluskey},\ and\
  \citenamefont {de~Walle}(2016)}]{Alkauskas2016}%
  \BibitemOpen
  \bibfield  {author} {\bibinfo {author} {\bibfnamefont {A.}~\bibnamefont
  {Alkauskas}}, \bibinfo {author} {\bibfnamefont {M.~D.}\ \bibnamefont
  {McCluskey}},\ and\ \bibinfo {author} {\bibfnamefont {C.~G.~V.}\ \bibnamefont
  {de~Walle}},\ }\bibfield  {title} {\enquote {\bibinfo {title} {Tutorial:
  Defects in semiconductors{\textemdash}combining experiment and theory},}\
  }\href {https://doi.org/10.1063/1.4948245} {\bibfield  {journal} {\bibinfo
  {journal} {Journal of Applied Physics}\ }\textbf {\bibinfo {volume} {119}},\
  \bibinfo {pages} {181101} (\bibinfo {year} {2016})}\BibitemShut {NoStop}%
\bibitem [{\citenamefont {Varley}\ \emph {et~al.}(2012)\citenamefont {Varley},
  \citenamefont {Janotti}, \citenamefont {Franchini},\ and\ \citenamefont
  {de~Walle}}]{Varley2012}%
  \BibitemOpen
  \bibfield  {author} {\bibinfo {author} {\bibfnamefont {J.~B.}\ \bibnamefont
  {Varley}}, \bibinfo {author} {\bibfnamefont {A.}~\bibnamefont {Janotti}},
  \bibinfo {author} {\bibfnamefont {C.}~\bibnamefont {Franchini}},\ and\
  \bibinfo {author} {\bibfnamefont {C.~G.~V.}\ \bibnamefont {de~Walle}},\
  }\bibfield  {title} {\enquote {\bibinfo {title} {Role of self-trapping in
  luminescence and p-type conductivity of wide-band-gap oxides},}\ }\href
  {https://doi.org/10.1103/physrevb.85.081109} {\bibfield  {journal} {\bibinfo
  {journal} {Physical Review B}\ }\textbf {\bibinfo {volume} {85}},\ \bibinfo
  {pages} {081109} (\bibinfo {year} {2012})}\BibitemShut {NoStop}%
\bibitem [{\citenamefont {Carey}\ \emph {et~al.}(1991)\citenamefont {Carey},
  \citenamefont {Spada}, \citenamefont {Berkowitz}, \citenamefont {Cao},\ and\
  \citenamefont {Thomas}}]{Carey1991}%
  \BibitemOpen
  \bibfield  {author} {\bibinfo {author} {\bibfnamefont {M.}~\bibnamefont
  {Carey}}, \bibinfo {author} {\bibfnamefont {F.}~\bibnamefont {Spada}},
  \bibinfo {author} {\bibfnamefont {A.}~\bibnamefont {Berkowitz}}, \bibinfo
  {author} {\bibfnamefont {W.}~\bibnamefont {Cao}},\ and\ \bibinfo {author}
  {\bibfnamefont {G.}~\bibnamefont {Thomas}},\ }\bibfield  {title} {\enquote
  {\bibinfo {title} {Preparation and structural characterization of sputtered
  {CoO}, {NiO}, and ni$_{0.5}$co$_{0.5}$o thin epitaxial films},}\ }\href
  {https://doi.org/10.1557/jmr.1991.2680} {\bibfield  {journal} {\bibinfo
  {journal} {Journal of Materials Research}\ }\textbf {\bibinfo {volume} {6}},\
  \bibinfo {pages} {2680--2687} (\bibinfo {year} {1991})}\BibitemShut {NoStop}%
\bibitem [{\citenamefont {Kurmaev}\ \emph {et~al.}(2008)\citenamefont
  {Kurmaev}, \citenamefont {Wilks}, \citenamefont {Moewes}, \citenamefont
  {Finkelstein}, \citenamefont {Shamin},\ and\ \citenamefont
  {Kune{\v{s}}}}]{Kurmaev2008}%
  \BibitemOpen
  \bibfield  {author} {\bibinfo {author} {\bibfnamefont {E.~Z.}\ \bibnamefont
  {Kurmaev}}, \bibinfo {author} {\bibfnamefont {R.~G.}\ \bibnamefont {Wilks}},
  \bibinfo {author} {\bibfnamefont {A.}~\bibnamefont {Moewes}}, \bibinfo
  {author} {\bibfnamefont {L.~D.}\ \bibnamefont {Finkelstein}}, \bibinfo
  {author} {\bibfnamefont {S.~N.}\ \bibnamefont {Shamin}},\ and\ \bibinfo
  {author} {\bibfnamefont {J.}~\bibnamefont {Kune{\v{s}}}},\ }\bibfield
  {title} {\enquote {\bibinfo {title} {Oxygen x-ray emission and absorption
  spectra as a probe of the electronic structure of strongly correlated
  oxides},}\ }\href {https://doi.org/10.1103/physrevb.77.165127} {\bibfield
  {journal} {\bibinfo  {journal} {Physical Review B}\ }\textbf {\bibinfo
  {volume} {77}},\ \bibinfo {pages} {165127} (\bibinfo {year}
  {2008})}\BibitemShut {NoStop}%
\bibitem [{\citenamefont {Sawatzky}\ and\ \citenamefont
  {Allen}(1984)}]{Sawatzky1984}%
  \BibitemOpen
  \bibfield  {author} {\bibinfo {author} {\bibfnamefont {G.~A.}\ \bibnamefont
  {Sawatzky}}\ and\ \bibinfo {author} {\bibfnamefont {J.~W.}\ \bibnamefont
  {Allen}},\ }\bibfield  {title} {\enquote {\bibinfo {title} {Magnitude and
  origin of the band gap in {NiO}},}\ }\href
  {https://doi.org/10.1103/physrevlett.53.2339} {\bibfield  {journal} {\bibinfo
   {journal} {Physical Review Letters}\ }\textbf {\bibinfo {volume} {53}},\
  \bibinfo {pages} {2339--2342} (\bibinfo {year} {1984})}\BibitemShut {NoStop}%
\bibitem [{\citenamefont {Powell}\ and\ \citenamefont
  {Spicer}(1970)}]{Powell1970}%
  \BibitemOpen
  \bibfield  {author} {\bibinfo {author} {\bibfnamefont {R.~J.}\ \bibnamefont
  {Powell}}\ and\ \bibinfo {author} {\bibfnamefont {W.~E.}\ \bibnamefont
  {Spicer}},\ }\bibfield  {title} {\enquote {\bibinfo {title} {Optical
  properties of {NiO} and {CoO}},}\ }\href
  {https://doi.org/10.1103/physrevb.2.2182} {\bibfield  {journal} {\bibinfo
  {journal} {Physical Review B}\ }\textbf {\bibinfo {volume} {2}},\ \bibinfo
  {pages} {2182--2193} (\bibinfo {year} {1970})}\BibitemShut {NoStop}%
\bibitem [{\citenamefont {Sokolov}\ \emph {et~al.}(2012)\citenamefont
  {Sokolov}, \citenamefont {Pustovarov}, \citenamefont {Churmanov},
  \citenamefont {Ivanov}, \citenamefont {Gruzdev}, \citenamefont {Sokolov},
  \citenamefont {Baranov},\ and\ \citenamefont {Moskvin}}]{Sokolov2012}%
  \BibitemOpen
  \bibfield  {author} {\bibinfo {author} {\bibfnamefont {V.~I.}\ \bibnamefont
  {Sokolov}}, \bibinfo {author} {\bibfnamefont {V.~A.}\ \bibnamefont
  {Pustovarov}}, \bibinfo {author} {\bibfnamefont {V.~N.}\ \bibnamefont
  {Churmanov}}, \bibinfo {author} {\bibfnamefont {V.~Y.}\ \bibnamefont
  {Ivanov}}, \bibinfo {author} {\bibfnamefont {N.~B.}\ \bibnamefont {Gruzdev}},
  \bibinfo {author} {\bibfnamefont {P.~S.}\ \bibnamefont {Sokolov}}, \bibinfo
  {author} {\bibfnamefont {A.~N.}\ \bibnamefont {Baranov}},\ and\ \bibinfo
  {author} {\bibfnamefont {A.~S.}\ \bibnamefont {Moskvin}},\ }\bibfield
  {title} {\enquote {\bibinfo {title} {Unusual x-ray excited luminescence
  spectra of {NiO} suggest self-trapping of the $d$-$d$ charge-transfer
  exciton},}\ }\href {https://doi.org/10.1103/physrevb.86.115128} {\bibfield
  {journal} {\bibinfo  {journal} {Physical Review B}\ }\textbf {\bibinfo
  {volume} {86}},\ \bibinfo {pages} {115128} (\bibinfo {year}
  {2012})}\BibitemShut {NoStop}%
\bibitem [{\citenamefont {Reinen}(1965)}]{Reinen1965}%
  \BibitemOpen
  \bibfield  {author} {\bibinfo {author} {\bibfnamefont {D.}~\bibnamefont
  {Reinen}},\ }\bibfield  {title} {\enquote {\bibinfo {title} {Farbe und
  {K}onstitution bei anorganischen {F}eststoffen. {VIII} [1] {B}. {D}ie
  {L}ichtabsorption des zweiwertigen {N}ickels in den {M}ischkristallen
  {Mg}$_{1-x}${Ni}$_x${O} und in tetraedrischer {K}oordination},}\ }\href
  {https://doi.org/10.1002/bbpc.19650690113} {\bibfield  {journal} {\bibinfo
  {journal} {Berichte der Bunsengesellschaft f\"ur physikalische Chemie}\
  }\textbf {\bibinfo {volume} {69}},\ \bibinfo {pages} {82--87} (\bibinfo
  {year} {1965})}\BibitemShut {NoStop}%
\bibitem [{\citenamefont {Kuiper}\ \emph {et~al.}(1989)\citenamefont {Kuiper},
  \citenamefont {Kruizinga}, \citenamefont {Ghijsen}, \citenamefont
  {Sawatzky},\ and\ \citenamefont {Verweij}}]{Kuiper1989}%
  \BibitemOpen
  \bibfield  {author} {\bibinfo {author} {\bibfnamefont {P.}~\bibnamefont
  {Kuiper}}, \bibinfo {author} {\bibfnamefont {G.}~\bibnamefont {Kruizinga}},
  \bibinfo {author} {\bibfnamefont {J.}~\bibnamefont {Ghijsen}}, \bibinfo
  {author} {\bibfnamefont {G.~A.}\ \bibnamefont {Sawatzky}},\ and\ \bibinfo
  {author} {\bibfnamefont {H.}~\bibnamefont {Verweij}},\ }\bibfield  {title}
  {\enquote {\bibinfo {title} {Character of holes in {Li}$_{x}${Ni$_{1-x}$}{O}
  and their magnetic behavior},}\ }\href
  {https://doi.org/10.1103/physrevlett.62.221} {\bibfield  {journal} {\bibinfo
  {journal} {Physical Review Letters}\ }\textbf {\bibinfo {volume} {62}},\
  \bibinfo {pages} {221--224} (\bibinfo {year} {1989})}\BibitemShut {NoStop}%
\bibitem [{\citenamefont {Reinert}\ \emph {et~al.}(1995)\citenamefont
  {Reinert}, \citenamefont {Steiner}, \citenamefont {H{\"u}fner}, \citenamefont
  {Schmitt}, \citenamefont {Fink}, \citenamefont {Knupfer}, \citenamefont
  {Sandl},\ and\ \citenamefont {Bertel}}]{Reinert1995}%
  \BibitemOpen
  \bibfield  {author} {\bibinfo {author} {\bibfnamefont {F.}~\bibnamefont
  {Reinert}}, \bibinfo {author} {\bibfnamefont {P.}~\bibnamefont {Steiner}},
  \bibinfo {author} {\bibfnamefont {S.}~\bibnamefont {H{\"u}fner}}, \bibinfo
  {author} {\bibfnamefont {H.}~\bibnamefont {Schmitt}}, \bibinfo {author}
  {\bibfnamefont {J.}~\bibnamefont {Fink}}, \bibinfo {author} {\bibfnamefont
  {M.}~\bibnamefont {Knupfer}}, \bibinfo {author} {\bibfnamefont
  {P.}~\bibnamefont {Sandl}},\ and\ \bibinfo {author} {\bibfnamefont
  {E.}~\bibnamefont {Bertel}},\ }\bibfield  {title} {\enquote {\bibinfo {title}
  {Electron and hole doping in {NiO}},}\ }\href
  {https://doi.org/10.1007/bf01317591} {\bibfield  {journal} {\bibinfo
  {journal} {Zeitschrift f\"{u}r Physik B Condensed Matter}\ }\textbf {\bibinfo
  {volume} {97}},\ \bibinfo {pages} {83--93} (\bibinfo {year}
  {1995})}\BibitemShut {NoStop}%
\bibitem [{\citenamefont {Mossanek}\ \emph {et~al.}(2013)\citenamefont
  {Mossanek}, \citenamefont {Dom{\'{\i}}nguez-Ca{\~{n}}izares}, \citenamefont
  {Guti{\'{e}}rrez}, \citenamefont {Abbate}, \citenamefont
  {D{\'{\i}}az-Fern{\'{a}}ndez},\ and\ \citenamefont {Soriano}}]{Mossanek2013}%
  \BibitemOpen
  \bibfield  {author} {\bibinfo {author} {\bibfnamefont {R.~J.~O.}\
  \bibnamefont {Mossanek}}, \bibinfo {author} {\bibfnamefont {G.}~\bibnamefont
  {Dom{\'{\i}}nguez-Ca{\~{n}}izares}}, \bibinfo {author} {\bibfnamefont
  {A.}~\bibnamefont {Guti{\'{e}}rrez}}, \bibinfo {author} {\bibfnamefont
  {M.}~\bibnamefont {Abbate}}, \bibinfo {author} {\bibfnamefont
  {D.}~\bibnamefont {D{\'{\i}}az-Fern{\'{a}}ndez}},\ and\ \bibinfo {author}
  {\bibfnamefont {L.}~\bibnamefont {Soriano}},\ }\bibfield  {title} {\enquote
  {\bibinfo {title} {Effects of {Ni} vacancies and crystallite size on the {O}
  1s and {Ni} 2p x-ray absorption spectra of nanocrystalline {NiO}},}\ }\href
  {https://doi.org/10.1088/0953-8984/25/49/495506} {\bibfield  {journal}
  {\bibinfo  {journal} {Journal of Physics: Condensed Matter}\ }\textbf
  {\bibinfo {volume} {25}},\ \bibinfo {pages} {495506} (\bibinfo {year}
  {2013})}\BibitemShut {NoStop}%
\bibitem [{\citenamefont {Ono}\ \emph {et~al.}(2018)\citenamefont {Ono},
  \citenamefont {Sasaki}, \citenamefont {Nagai}, \citenamefont {Yamaguchi},
  \citenamefont {Higashiwaki}, \citenamefont {Kuramata}, \citenamefont
  {Yamakoshi}, \citenamefont {Sato}, \citenamefont {Honda},\ and\ \citenamefont
  {Onuma}}]{Ono2018}%
  \BibitemOpen
  \bibfield  {author} {\bibinfo {author} {\bibfnamefont {M.}~\bibnamefont
  {Ono}}, \bibinfo {author} {\bibfnamefont {K.}~\bibnamefont {Sasaki}},
  \bibinfo {author} {\bibfnamefont {H.}~\bibnamefont {Nagai}}, \bibinfo
  {author} {\bibfnamefont {T.}~\bibnamefont {Yamaguchi}}, \bibinfo {author}
  {\bibfnamefont {M.}~\bibnamefont {Higashiwaki}}, \bibinfo {author}
  {\bibfnamefont {A.}~\bibnamefont {Kuramata}}, \bibinfo {author}
  {\bibfnamefont {S.}~\bibnamefont {Yamakoshi}}, \bibinfo {author}
  {\bibfnamefont {M.}~\bibnamefont {Sato}}, \bibinfo {author} {\bibfnamefont
  {T.}~\bibnamefont {Honda}},\ and\ \bibinfo {author} {\bibfnamefont
  {T.}~\bibnamefont {Onuma}},\ }\bibfield  {title} {\enquote {\bibinfo {title}
  {Relation between electrical and optical properties of p-type {NiO} films},}\
  }\href {https://doi.org/10.1002/pssb.201700311} {\bibfield  {journal}
  {\bibinfo  {journal} {physica status solidi (b)}\ }\textbf {\bibinfo {volume}
  {255}},\ \bibinfo {pages} {1700311} (\bibinfo {year} {2018})}\BibitemShut
  {NoStop}%
\bibitem [{\citenamefont {Allen}\ and\ \citenamefont {Dyke}(1976)}]{Allen1976}%
  \BibitemOpen
  \bibfield  {author} {\bibinfo {author} {\bibfnamefont {G.}~\bibnamefont
  {Allen}}\ and\ \bibinfo {author} {\bibfnamefont {J.}~\bibnamefont {Dyke}},\
  }\bibfield  {title} {\enquote {\bibinfo {title} {An investigation of the
  optical spectrum of lithium doped nickel oxide},}\ }\href
  {https://doi.org/10.1016/0009-2614(76)80240-0} {\bibfield  {journal}
  {\bibinfo  {journal} {Chemical Physics Letters}\ }\textbf {\bibinfo {volume}
  {37}},\ \bibinfo {pages} {391--395} (\bibinfo {year} {1976})}\BibitemShut
  {NoStop}%
\bibitem [{\citenamefont {Karsthof}\ \emph
  {et~al.}(2020{\natexlab{b}})\citenamefont {Karsthof}, \citenamefont {Anton},
  \citenamefont {Kremer},\ and\ \citenamefont {Grundmann}}]{Karsthof2020a}%
  \BibitemOpen
  \bibfield  {author} {\bibinfo {author} {\bibfnamefont {R.}~\bibnamefont
  {Karsthof}}, \bibinfo {author} {\bibfnamefont {A.~M.}\ \bibnamefont {Anton}},
  \bibinfo {author} {\bibfnamefont {F.}~\bibnamefont {Kremer}},\ and\ \bibinfo
  {author} {\bibfnamefont {M.}~\bibnamefont {Grundmann}},\ }\bibfield  {title}
  {\enquote {\bibinfo {title} {Nickel vacancy acceptor in nickel oxide: Doping
  beyond thermodynamic equilibrium},}\ }\href
  {https://doi.org/10.1103/physrevmaterials.4.034601} {\bibfield  {journal}
  {\bibinfo  {journal} {Physical Review Materials}\ }\textbf {\bibinfo {volume}
  {4}},\ \bibinfo {pages} {034601} (\bibinfo {year}
  {2020}{\natexlab{b}})}\BibitemShut {NoStop}%
\bibitem [{\citenamefont {Guti\'{e}rrez}\ \emph {et~al.}(2020)\citenamefont
  {Guti\'{e}rrez}, \citenamefont {Dom\'{\i}nguez-Ca{\~{n}}izares},
  \citenamefont {Krause}, \citenamefont {D\'{\i}az-Fern\'{a}ndez},\ and\
  \citenamefont {Soriano}}]{Gutierrez2020}%
  \BibitemOpen
  \bibfield  {author} {\bibinfo {author} {\bibfnamefont {A.}~\bibnamefont
  {Guti\'{e}rrez}}, \bibinfo {author} {\bibfnamefont {G.}~\bibnamefont
  {Dom\'{\i}nguez-Ca{\~{n}}izares}}, \bibinfo {author} {\bibfnamefont
  {S.}~\bibnamefont {Krause}}, \bibinfo {author} {\bibfnamefont
  {D.}~\bibnamefont {D\'{\i}az-Fern\'{a}ndez}},\ and\ \bibinfo {author}
  {\bibfnamefont {L.}~\bibnamefont {Soriano}},\ }\bibfield  {title} {\enquote
  {\bibinfo {title} {Thermal induced depletion of cationic vacancies in {NiO}
  thin films evidenced by x-ray absorption spectroscopy at the {O} 1s
  threshold},}\ }\href {https://doi.org/10.1116/6.0000080} {\bibfield
  {journal} {\bibinfo  {journal} {Journal of Vacuum Science {\&} Technology A}\
  }\textbf {\bibinfo {volume} {38}},\ \bibinfo {pages} {033209} (\bibinfo
  {year} {2020})}\BibitemShut {NoStop}%
\bibitem [{\citenamefont {D{\'i}az-Guerra}\ \emph {et~al.}(1997)\citenamefont
  {D{\'i}az-Guerra}, \citenamefont {Rem{\'{o}}n}, \citenamefont {Garc{\'i}a},\
  and\ \citenamefont {Piqueras}}]{DiazGuerra1997}%
  \BibitemOpen
  \bibfield  {author} {\bibinfo {author} {\bibfnamefont {C.}~\bibnamefont
  {D{\'i}az-Guerra}}, \bibinfo {author} {\bibfnamefont {A.}~\bibnamefont
  {Rem{\'{o}}n}}, \bibinfo {author} {\bibfnamefont {J.~A.}\ \bibnamefont
  {Garc{\'i}a}},\ and\ \bibinfo {author} {\bibfnamefont {J.}~\bibnamefont
  {Piqueras}},\ }\bibfield  {title} {\enquote {\bibinfo {title}
  {Cathodoluminescence and photoluminescence spectroscopy of {NiO}},}\ }\href
  {https://doi.org/10.1002/1521-396x(199710)163:2<497::aid-pssa497>3.0.co;2-z}
  {\bibfield  {journal} {\bibinfo  {journal} {physica status solidi (a)}\
  }\textbf {\bibinfo {volume} {163}},\ \bibinfo {pages} {497--503} (\bibinfo
  {year} {1997})}\BibitemShut {NoStop}%
\bibitem [{\citenamefont {Volkov}, \citenamefont {Wang},\ and\ \citenamefont
  {Zou}(2001)}]{Volkov2001}%
  \BibitemOpen
  \bibfield  {author} {\bibinfo {author} {\bibfnamefont {V.~V.}\ \bibnamefont
  {Volkov}}, \bibinfo {author} {\bibfnamefont {Z.}~\bibnamefont {Wang}},\ and\
  \bibinfo {author} {\bibfnamefont {B.}~\bibnamefont {Zou}},\ }\bibfield
  {title} {\enquote {\bibinfo {title} {Carrier recombination in clusters of
  {NiO}},}\ }\href {https://doi.org/10.1016/s0009-2614(01)00191-9} {\bibfield
  {journal} {\bibinfo  {journal} {Chemical Physics Letters}\ }\textbf {\bibinfo
  {volume} {337}},\ \bibinfo {pages} {117--124} (\bibinfo {year}
  {2001})}\BibitemShut {NoStop}%
\bibitem [{\citenamefont {Ho}\ \emph {et~al.}(2015)\citenamefont {Ho},
  \citenamefont {Kuo}, \citenamefont {Chan},\ and\ \citenamefont
  {Ma}}]{Ho2015}%
  \BibitemOpen
  \bibfield  {author} {\bibinfo {author} {\bibfnamefont {C.-H.}\ \bibnamefont
  {Ho}}, \bibinfo {author} {\bibfnamefont {Y.-M.}\ \bibnamefont {Kuo}},
  \bibinfo {author} {\bibfnamefont {C.-H.}\ \bibnamefont {Chan}},\ and\
  \bibinfo {author} {\bibfnamefont {Y.-R.}\ \bibnamefont {Ma}},\ }\bibfield
  {title} {\enquote {\bibinfo {title} {Optical characterization of strong {UV}
  luminescence emitted from the excitonic edge of nickel oxide nanotowers},}\
  }\href {https://doi.org/10.1038/srep15856} {\bibfield  {journal} {\bibinfo
  {journal} {Scientific Reports}\ }\textbf {\bibinfo {volume} {5}},\ \bibinfo
  {pages} {15856} (\bibinfo {year} {2015})}\BibitemShut {NoStop}%
\bibitem [{\citenamefont {Dawson}, \citenamefont {Guo},\ and\ \citenamefont
  {Robertson}(2015)}]{Dawson2015}%
  \BibitemOpen
  \bibfield  {author} {\bibinfo {author} {\bibfnamefont {J.~A.}\ \bibnamefont
  {Dawson}}, \bibinfo {author} {\bibfnamefont {Y.}~\bibnamefont {Guo}},\ and\
  \bibinfo {author} {\bibfnamefont {J.}~\bibnamefont {Robertson}},\ }\bibfield
  {title} {\enquote {\bibinfo {title} {Energetics of intrinsic defects in {NiO}
  and the consequences for its resistive random access memory performance},}\
  }\href {https://doi.org/10.1063/1.4931751} {\bibfield  {journal} {\bibinfo
  {journal} {Applied Physics Letters}\ }\textbf {\bibinfo {volume} {107}},\
  \bibinfo {pages} {122110} (\bibinfo {year} {2015})}\BibitemShut {NoStop}%
\bibitem [{\citenamefont {Varley}\ \emph {et~al.}(2010)\citenamefont {Varley},
  \citenamefont {Weber}, \citenamefont {Janotti},\ and\ \citenamefont
  {de~Walle}}]{Varley2010}%
  \BibitemOpen
  \bibfield  {author} {\bibinfo {author} {\bibfnamefont {J.~B.}\ \bibnamefont
  {Varley}}, \bibinfo {author} {\bibfnamefont {J.~R.}\ \bibnamefont {Weber}},
  \bibinfo {author} {\bibfnamefont {A.}~\bibnamefont {Janotti}},\ and\ \bibinfo
  {author} {\bibfnamefont {C.~G.~V.}\ \bibnamefont {de~Walle}},\ }\bibfield
  {title} {\enquote {\bibinfo {title} {Oxygen vacancies and donor impurities in
  $beta$-ga2o3},}\ }\href {https://doi.org/10.1063/1.3499306} {\bibfield
  {journal} {\bibinfo  {journal} {Applied Physics Letters}\ }\textbf {\bibinfo
  {volume} {97}},\ \bibinfo {pages} {142106} (\bibinfo {year}
  {2010})}\BibitemShut {NoStop}%
\bibitem [{\citenamefont {Frodason}\ \emph {et~al.}()\citenamefont {Frodason},
  \citenamefont {Johansen}, \citenamefont {Galeckas},\ and\ \citenamefont
  {Vines}}]{Frodason2019}%
  \BibitemOpen
  \bibfield  {author} {\bibinfo {author} {\bibfnamefont {Y.~K.}\ \bibnamefont
  {Frodason}}, \bibinfo {author} {\bibfnamefont {K.~M.}\ \bibnamefont
  {Johansen}}, \bibinfo {author} {\bibfnamefont {A.}~\bibnamefont {Galeckas}},\
  and\ \bibinfo {author} {\bibfnamefont {L.}~\bibnamefont {Vines}},\ }\bibfield
   {title} {\enquote {\bibinfo {title} {Broad luminescence from donor-complexed
  {Li$_{\mathrm{Zn}}$} and {Na$_{\mathrm{Zn}}$} acceptors in {ZnO}},}\ }\href
  {https://doi.org/10.1103/physrevb.100.184102} {\ \textbf {\bibinfo {volume}
  {100}},\ \bibinfo {pages} {184102}}\BibitemShut {NoStop}%
\end{thebibliography}
\end{document}